# Factoring sparse polynomials fast*†

ALEXANDER DEMIN
National Research University Higher School of Economics
20 Myasnitskaya Ulitsa
101000 Moscow, Russia
*Email:* asdemin_2@edu.hse.ru

JORIS VAN DER HOEVEN
CNRS, École polytechnique, Institut Polytechnique de Paris
Laboratoire d'informatique de l'École polytechnique (LIX, UMR 7161)
Bâtiment Alan Turing, CS35003
1, rue Honoré d'Estienne d'Orves
91120 Palaiseau, France
*Email:* vdhoeven@lix.polytechnique.fr

Consider a sparse polynomial in several variables given explicitly as a sum of non-zero terms with coefficients in an effective field. In this paper, we present several algorithms for factoring such polynomials and related tasks (such as gcd computation, square-free factorization, content-free factorization, and root extraction). Our methods are all based on sparse interpolation, but follow two main lines of attack: iteration on the number of variables and more direct reductions to the univariate or bivariate case. We present detailed probabilistic complexity bounds in terms of the complexity of sparse interpolation and evaluation.

## 1. INTRODUCTION

### 1.1. Motivation and main goals

Let $\mathbb{K}$ be an effective field. Consider a sparse polynomial $F \in \mathbb{K}[x_1, \ldots, x_n]$, represented as

$$F = F_1 x^{\gamma_1} + \cdots + F_s x^{\gamma_s}, \tag{1.1}$$

where $F_1, \ldots, F_s \in \mathbb{K}^{\neq} := \mathbb{K} \setminus \{0\}$, $\gamma_1, \ldots, \gamma_s \in \mathbb{N}^n$, and $x^e := x_1^{e_1} \cdots x_n^{e_n}$ for any $e = (e_1, \ldots, e_n) \in \mathbb{N}^n$. We call $s_F := s$ the *size* of $F$ and $\operatorname{supp} F := \{\gamma_1, \ldots, \gamma_s\}$ its *support*. The aim of this paper is to factor $F$ into a product of irreducible sparse polynomials.

All algorithms that we will present are based on the approach of *sparse evaluation* and *interpolation*. Instead of directly working with sparse representations (1.1), the idea is to evaluate input polynomials at a sequence of well-chosen points, do the actual work on these evaluations, and then recover the output polynomials using sparse interpolation. The evaluation-interpolation approach leads to very efficient algorithms for many tasks, such as multiplication [48, 46], division, gcd computations [51], etc. In this paper, we investigate the complexity of factorization under this light.

*. This work has partly been supported by the French ANR-22-CE48-0016 NODE project.

†. This article has been written using GNU $\mathrm{T_EX_{MACS}}$ [47].

One particularly good way to choose the evaluation points is to take them in a geometric progression: for a fixed $\alpha = (\alpha_1, \ldots, \alpha_n) \in (\mathbb{K}^{\neq})^n$, we evaluate at $\alpha, \alpha^2, \alpha^3, \ldots$, where $\alpha^k := (\alpha_1^k, \ldots, \alpha_n^k)$. This idea goes back to Prony [91] and was rediscovered, extended, and popularized by Ben Or and Tiwari [5]. We refer to [92] for a nice survey. If $\mathbb{K}$ is a finite field, then a further refinement is to use suitable roots of unity, in which case both sparse evaluation and interpolation essentially reduce to discrete Fourier transforms [52, 46].

In this paper, we do not specify the precise algorithms that should be used for sparse evaluation and interpolation, but we will always assume that the evaluation points form geometric progressions. Then the cost $\mathsf{S}(s)$ of sparse evaluation or interpolation at $s$ such points is quasi-linear in $s$. We refer to Sections 2.1, 2.3, and 2.4 for more details on this cost function $\mathsf{S}(s)$ and the algebraic complexity model that we assume.

One important consequence of relying on geometric progressions is that this constraints the type of factorization algorithms that will be efficient. For instance, several existing methods start with the application of random shifts $x_i \longmapsto x_i + \sigma_i$ for one or more variables $x_i$. Since such shifts do not preserve geometric progressions, this is a technique that we must avoid. On the other hand, monomial transformations like $x_i \longmapsto y_1^{w_{i,1}} \cdots y_n^{w_{i,n}}$ do preserve geometric progressions and we will see how to make use of this fact.

The main goal of this paper is to develop fast algorithms for factoring sparse polynomials under these constraints. Besides the top-level problem of factorization into irreducibles, we also consider several interesting subtasks, such as gcd computations, Hensel lifting, content-free and square-free factorization, and the extraction of multiple roots. While relying on known techniques, we shall show how to conciliate them with the above constraints.

Our complexity bounds are expressed in terms of the maximal size and total degree of the input and output polynomials. In practical applications, total degrees often remain reasonably small, so we typically allow for a polynomial dependence on the total degree times the required number of evaluation/interpolation points. In this particular asymptotic regime, our complexity bounds are very sharp and they improve on the bounds from the existing literature.

Concerning the top-level problem of decomposing sparse polynomials into irreducible factors, we develop two main approaches: a recursive one on the dimension and a more direct one based on simultaneous lifting with respect to all but one variables. We will present precise complexity bounds and examples of particularly difficult cases.

## 1.2. Overview of univariate and bivariate factorization methods

The factorization of polynomials is a fundamental problem in computer algebra. Since we are relying on sparse interpolation techniques, it is also natural to focus exclusively on randomized algorithms of Monte Carlo type. For some deterministic algorithms, we refer to [61, 101, 72].

Before considering multivariate polynomials, we need an algorithm for factoring univariate polynomials. Throughout this paper, we assume that we have an algorithm for this task (it can be shown that the mere assumption of $\mathbb{K}$ being effective is not sufficient [27, 28]).

In practice, we typically have $\mathbb{K} = \mathbb{F}_{p^\kappa}$, $\mathbb{K} = \mathbb{Q}$, $\mathbb{K} \subseteq \mathbb{Q}_p$, or $\mathbb{K} \subseteq \mathbb{C}$ for some prime $p$ and $\kappa \geqslant 1$. Most basic is the case when $\mathbb{K}$ is a finite field, and various efficient probabilistic methods have been proposed for this case. An early such method is due to Berlekamp [6, 7]. A very efficient algorithm that is also convenient to implement is due to Cantor and Zassenhaus [16]. Asymptotically more efficient methods have been developed since [67, 70] as well as specific improvements for the case when $\kappa$ is composite [53]. See also [31, Chapter 14] and [65].

Rational numbers can either be regarded as a subfield of $\mathbb{C}$ or $\mathbb{Q}_p$. For asymptotically efficient algorithms for the approximate factorizations of univariate polynomials over $\mathbb{C}$, we refer to [98, 87, 83]. When reducing a polynomial in $\mathbb{Q}$ modulo $p$ for a sufficiently large random prime, factorization over $\mathbb{Q}_p$ reduces to factorization over $\mathbb{F}_p$ via Hensel lifting [97, 42, 107]. For more general factorization methods over $\mathbb{Q}_p$, we refer to [26, 20, 82, 39, 3].

Now let $F \in \mathbb{Q}[x]$ and assume that we have an irreducible factorization $F = P_1 \cdots P_\ell$ with $P_1,\ldots,P_\ell \in \mathbb{K}[x]$ for $\mathbb{K} \subseteq \mathbb{C}$ or $\mathbb{K} \subseteq \mathbb{Q}_p$. (In practice, we require that $P_1,\ldots,P_\ell$ are known with sufficient precision.) In order to turn this into a factorization over $\mathbb{Q}$, we need a way to recombine the factors, i.e., to find the subsets $I \subseteq \{1, \ldots, \ell\}$ for which $\prod_{i \in I} P_i \in \mathbb{Q}[x]$. Indeed, if $F$ is irreducible in $\mathbb{Q}[x]$ and $d := \deg F \geqslant 2$, then $F$ is never irreducible in $\mathbb{C}[x]$ and irreducible with probability $\approx 1/d$ in $\mathbb{F}_p[x]$ for a large random prime $p$. The first recombination method that runs in polynomial time is due to Lenstra-Lenstra-Lovasz [77]. Subsequently, many improvements and variants of this LLL-algorithm have been developed [44, 4, 85, 86, 94].

The problem of factoring a bivariate polynomial $F \in \mathbb{K}[x, y]$ over $\mathbb{K}$ is similar in many regards to factoring polynomials with rational coefficients. Indeed, for a suitable random prime $p$, we have seen above that the latter problem reduces to univariate factorization over $\mathbb{F}_p$, Hensel lifting, and factor combination. In a similar way, after factoring $F(x, \tau)$ for a sufficiently random $\tau$ (possibly in an extension field of $\mathbb{K}$, whenever $\mathbb{K}$ is a small finite field), we may use Hensel lifting to obtain a factorization in $\mathbb{K}[[y - \tau]][x]$, and finally recombine the factors. The precise algorithms for factor recombination are slightly different in this context [29, 71, 4, 74] (see also [93, 95] for earlier related work), although they rely on similar ideas.

## 1.3. Overview of multivariate factorization methods

Hensel lifting naturally generalizes to polynomials in three or more variables $x_1, \ldots, x_n$. Many algorithms for multivariate polynomial factorization rely on it [84, 104, 103, 109, 60, 32, 33, 62, 61, 8, 72, 80, 81, 17], as well as many implementations in computer algebra systems. One important property of higher dimensional Hensel lifting is that factor recombination can generally be avoided with high probability, contrary to what we saw for $p$-adic and bivariate Hensel lifting. This is due to Hilbert and Bertini's irreducibility theorems [43, 10]:

THEOREM 1.1. *Assume that $F \in \mathbb{K}[x_1, \ldots, x_n] \setminus \mathbb{K}$ is irreducible and of total degree $d$. Let $U$ be the set of points $(\alpha_1, \ldots, \alpha_n, \beta_1, \ldots, \beta_n, \gamma_1, \ldots, \gamma_n) \in \mathbb{K}^{3n}$ for which*

$$F(\alpha_1 t + \beta_1 u + \gamma_1, \ldots, \alpha_n t + \beta_n u + \gamma_n) \qquad (1.2)$$

*is irreducible in $\mathbb{K}[t, u]$. Then $U$ is a Zariski open subset of $\mathbb{K}^{3n}$, which is dense over the algebraic closure of $\mathbb{K}$.*

Effective versions of these theorems can be used to directly reduce the factorization problem in dimension $n$ to a bivariate problem (and even to a univariate problem if $\mathbb{K} = \mathbb{Q}$, using similar ideas). We refer to [74] and [75, Chapître 6] for a recent presentation of how to do this and to [100, Section 6.1] for the mathematical background.

In order to analyze the computational complexity of factorization, we first need to specify the precise way we represent our polynomials. When using a dense representation (e.g. storing all monomials of total degree $\leqslant d$ with their (possibly zero) coefficients), Theorem 1.1 allows us to directly reduce to the bivariate case (if $\mathbb{K}$ is very small, then one may need to replace $\mathbb{K}$ by a suitable algebraic extension). The first polynomial time reduction of this kind was proposed by Kaltofen [60]. More recent bounds that exploit fast dense polynomial arithmetic can be found in [72].

Another popular representation is the "black box representation", in which case we are only given an algorithm for the evaluation of our polynomial $F$. Often this algorithm is actually a straight line program (SLP) [14], which corresponds to the "SLP representation". In these cases, the relevant complexity measure is the length of the SLP or the maximal number of steps that are needed to compute one black box evaluation. Consequently, affine changes of variables (1.2) only marginally increase the program size. This has been exploited in order to derive polynomial time complexity bounds for factoring in this model [63, 68]; see also [30, 17, 18]. Here we stress that the output factors are also given in black box or SLP representation.

Plausibly the most common representation for multivariate polynomials in computer algebra is the sparse one (1.1). Any sparse polynomial naturally gives rise to an SLP of roughly the same size. Sparse interpolation also provides a way to convert in the opposite direction. However, for an SLP $F^{\mathrm{slp}}$ of length $L$, it takes $\Theta(Ls)$ time to recover its sparse representation $F$, where $s := s_F$. *A priori*, the back and forth conversion $F \mapsto F^{\mathrm{slp}} \mapsto F$ therefore takes quadratic time $\Theta(s^2)$. While it is theoretically possible to factor sparse polynomials using the above black box methods, this is suboptimal from a complexity perspective.

Unfortunately, general affine transformations (1.2) destroy sparsity, so additional ideas are needed for the design of efficient factorization methods based on Hilbert-Bertini-like irreducibility theorems. Dedicated algorithms for the sparse model have been developed in [32, 8, 79, 81]. There are two ways to counter the loss of sparsity under affine transformations, both of which will be considered in the present paper. One possibility is to successively use Hensel lifting with respect to $x_3, \ldots, x_n$. Another approach is to use a more particular type of changes of variables, like $F(t, \alpha_2 u, \ldots, \alpha_n u)$. However, both approaches require $F$ to be of a suitable form to allow for Hensel lifting. Some reference for Hensel lifting in the context of sparse polynomials are [62, 8, 78, 80, 81, 17].

For most applications in computer algebra, the total degree of large sparse polynomials is typically much smaller than the number of terms. The works in the above references mainly target this asymptotic regime. The factorization of "supersparse" polynomials has also been considered in [40, 37]. The survey talk [96] discusses the complexity of polynomial factorizations for yet other polynomial representations.

The theory of polynomial factorization involves several other basic algorithmic tasks that are interesting for their own sake. We already mentioned the importance of Hensel lifting. Other fundamental operations are gcd computations, multiple root extractions, square-free factorizations, and determining the content of a polynomial. We refer to [31] for classical univariate algorithms. As to sparse multivariate polynomials, there exist many approaches for gcd computations [21, 108, 68, 66, 59, 22, 56, 76, 51, 58].

Whenever convenient, we will assume that the characteristic of $\mathbb{K}$ is either zero or sufficiently large. This will allow us to avoid technical non-separability issues; we refer to [34, 102, 73] for algorithms to deal with such issues. A survey of multivariate polynomial factorization over finite fields (including small characteristic) can be found in [65].

Throughout this paper, factorizations will be done over the ground field $\mathbb{K}$. Some algorithms for "absolute factorization" over the algebraic closure $\mathbb{K}^{\mathrm{alg}}$ of $\mathbb{K}$ can be found in [64, 19, 23]; alternatively, one may mimic computations in $\mathbb{K}^{\mathrm{alg}}$ using dynamic evaluation [24, 49].

## 1.4. Outline of our contributions

The goal of this paper is to redevelop the theory of sparse polynomial factorization, by taking advantage as fully as possible of evaluation-interpolation techniques at geometric sequences. The basic background material is recalled in Section 2. We recall that almost all algorithms in the paper are randomized, of Monte Carlo type. We also note that the correctness of a factorization can easily be verified, either directly or with high probability, by evaluating the polynomial and the product of the potential factors at a random point. (On the other hand, this approach does not yield a probabilistic irreducibility test.)

As an appetizer, we start in Section 3 with the problem of multivariate gcd computation. This provides a nice introduction for the main two approaches used later in the paper: induction on dimension and direct reduction to the univariate (or bivariate or low dimensional) case. It also illustrates the kind of complexity bounds that we are aiming for. Consider the computation of the gcd $G$ of two sparse polynomials $P,Q \in \mathbb{K}[x_1,\ldots,x_n]$. Ideally speaking, setting $s := s_G$, $\bar{s} := \max(s_P, s_Q, s_G)$, $d := \max(d_P, d_Q)$, we are aiming for a complexity bound of the form

$$\tilde{O}(\bar{s} + d^{\vartheta} s), \tag{1.3}$$

where $\vartheta$ is a constant. Since $s$ is typically much smaller than $\bar{s}$, we can afford ourselves the non-trivial overhead with respect to $d$ in the term $d^{\vartheta} s$. The inductive approach on the dimension $n$ achieves the complexity (1.3) with $\vartheta = 1$, but its worst case complexity is $\tilde{O}(n(\bar{s} + d s))$. This algorithm seems to be new. The second approach uses a direct reduction to univariate gcd computations via so-called "regularizing weights" and achieves the complexity (1.3) with $\vartheta \leqslant 2$, and even $\vartheta = 1$ for some practical examples (e.g., when the gcd is monic in one of the variables). This algorithm is a sharpened version of the algorithm from [51, Section 4.3].

Most existing algorithms for multivariate polynomial factorization reduce to the univariate or bivariate case. Direct reduction to the univariate case is only possible for certain types of coefficients, such as integers, rational numbers, or algebraic numbers. Reduction to the bivariate case works in general, thanks to Theorem 1.1, and this is even interesting when $\mathbb{K} = \mathbb{Q}$: first reduce modulo a suitable prime $p$, then factor over $\mathbb{F}_p$, and finally Hensel lift to obtain a factorization over $\mathbb{Q}$. In this paper, we will systematically opt for the bivariate reduction strategy. For this reason, we have included a separate Section 4 on bivariate factorization and related problems. This material is classical, but recalled for self-containedness and convenience of the reader.

If $\mathbb{K}$ is a finite field, then we already noted that multivariate factorization does not directly reduce to the univariate case. Nevertheless, such direct reductions are possible for some special cases of interest: content-free factorization, extraction of multiple roots, and square-free factorization. In Section 5, we present efficient algorithms for these tasks, following the "regularizing weights" approach that was introduced in Section 3.2 for gcd computations. All complexity bounds are of the form (1.3) for the relevant notions of output size $s$, input-output size $\bar{s}$, and total degree $d$.

In Section 6, we turn to the factorization of a multivariate polynomial $F \in \mathbb{K}[x_1, \ldots, x_n]$ using induction on the dimension $n$. Starting with a coprime factorization of a bivariate projection $F(x_1, x_2, c_3, \ldots, c_n) \in \mathbb{K}[x_1, x_2]$ of $F$ for random $c_3, \ldots, c_n$, we successively Hensel lift this factorization with respect to $x_3, \ldots, x_n$. Using a suitable evaluation-interpolation strategy, the actual Hensel lifting can be done on bivariate polynomials. This leads to complexity bounds of the form $\tilde{O}(n(\bar{s} + d^{\vartheta} s))$ with $\vartheta = 2$. In fact, we separately consider factorizations into two or more factors. In the case of two factors, it is possible to first determine the smallest factor and then perform an exact division to obtain the other one. In the complexity bound, this means that $s$ should really be understood as the number of evaluation-interpolation points, i.e. the minimum of the sizes of the two factors. It depends on the situation whether it is faster to lift a full coprime factorization or one factor at a time, although we expect the first approach to be fastest in most cases.

Due to the fact that projections such as $F(x_1, x_2, c_3, \ldots, c_n)$ are only very special types of affine transformations, Theorem 1.1 does not apply. Therefore, the direct inductive approach from Section 6 does not systematically lead to a full irreducible factorization of $F$. In Remarks 6.14 and 6.15, we give an example where our approach fails, together with two different heuristic remedies (which both lead to similar complexity bounds, but with $\theta = 3$ or higher).

In the last Section 7, we present another approach, which avoids induction on the dimension $n$ and the corresponding overhead in the complexity bound. The idea is to exploit properties of the Newton polytope of $F$ and "face factorizations" (e.g. factorizations of restrictions of $F$ to faces of its Newton polytope). In the most favorable case, there exists a coprime edge factorization, which can be Hensel lifted into a full factorization, and we obtain a complexity bound of the form (1.3). In less favorable cases, different face factorizations need to be combined. Although this yields a similar complexity bound, the details are more technical. We refer to [1, 105] for a few existing ways to exploit Newton polytopes for polynomial factorization.

In very unfavorable cases (see Section 7.6), the factorization of $F$ cannot be recovered from its face factorizations at all. In Section 7.7, we conclude with a fully general algorithm for irreducible factorization. This algorithm is not as practical, but addresses pathologically difficult cases through the introduction of a few extra variables. Its cost is $\tilde{O}(d^3 \bar{s} + d^{10})$ plus the cost of one univariate factorization of degree $O(d^2)$.

## 1.5. Notation

In this paper, we will measure the complexity of algorithms using the algebraic complexity model [14]. In addition, we assume that it is possible to sample a random element from $\mathbb{K}$ (or a subset of $\mathbb{K}$) in constant time. We will use $\tilde{O}(g(n))$ as a shorthand for $g(n)(\log g(n))^{O(1)}$.

We let $\mathbb{N} := \{0, 1, 2, \ldots\}$ and $\mathbb{N}^{>} := \{1, 2, 3, \ldots\}$. We also define $R^{\neq} := \{x \in R : x \neq 0\}$, for any ring $R$. Given a multivariate polynomial $F \in \mathbb{K}[x_1, \ldots, x_n]$ and $i \in \{1, \ldots, n\}$, we write $\delta_i := \deg_{x_i} F$ (resp. $\mathrm{val}_{x_i} F$) for the degree (resp. valuation) of $F$ in $x_i$. We also write $d_F := \deg F$ (resp. $\mathrm{val}\, F$) for the total degree (resp. valuation of $F$), and we set $\delta_F := \max_i \deg_{x_i} F$. Recall that $s_F$ stands for the number of terms of $F$ in its sparse representation (1.1).

**Acknowledgment.** We wish to thank Grégoire Lecerf for useful discussions during the preparation of this paper and the first anonymous referee for helpful comments and suggestions.

# 2. PRELIMINARIES

## 2.1. Basic complexities

Let $\mathsf{M}(n)$ (or $\mathsf{M}_{\mathbb{K}}(n)$) be the cost to multiply two dense univariate polynomials of degree $\leqslant n$ in $\mathbb{K}[x]$. Throughout the paper, we make the assumption that $\mathsf{M}(n)/n$ is a non-decreasing function. In the algebraic complexity model [14], when counting the number of operations in $\mathbb{K}$, we may take $\mathsf{M}(n) = O(n \log n \log \log n)$ [15]. If $\mathbb{K}$ is a finite field $\mathbb{F}_q$, then one has $\mathsf{M}_{\mathbb{F}_q}(n) = O(n \log n)$, under suitable number theoretic assumption [41]. In this case, the corresponding bit complexity (when counting the number of operations on a Turing machine [88]) is $O(n \log q \log (n \log q))$.

For polynomials $f, g \in \mathbb{K}[x]^{\neq}$ of degree $\leqslant n$ it is possible to find the unique $q, r \in \mathbb{K}[x]$, such that $f = qg + r$ with $\deg r < n$. This is the problem of *univariate division with remainder*, which can be solved in time $O(\mathsf{M}(n))$ by applying Newton iteration [31, Section 9.1]. A related task is *univariate root extraction*: given $f \in \mathbb{K}[x]$ and $\ell \in \mathbb{N}^{>}$, check whether $f$ is of the form $f = c g^{\ell}$ for some $c \in \mathbb{K}^{\neq}$ and monic $g \in \mathbb{K}[x]$, and determine $c$ and $g$ if so. For a fixed $\ell$, this can be checked, and the unique $g$ can be found in $O(\mathsf{M}(n))$ arithmetic operations in $\mathbb{K}$ by applying Newton iteration [31, polynomial analogue of Theorem 9.28].

Consider the Vandermonde matrix

$$V = \begin{pmatrix} 1 & \alpha_1 & \cdots & \alpha_1^{n-1} \\ \vdots & & & \vdots \\ 1 & \alpha_n & \cdots & \alpha_n^{n-1} \end{pmatrix},$$

where $\alpha_1, \ldots, \alpha_n \in \mathbb{K}$ are pairwise distinct. Given a column vector $C$ with entries $c_0, \ldots, c_{n-1}$, it is well known [11, 69, 12, 9, 45] that the products $VC$, $V^{-1}C$, $V^{\top}C$, and $(V^{-1})^{\top}C$ can all be computed in time $O(\mathsf{M}(n) \log n)$. These are respectively the problems of *multi-point evaluation*, (*univariate*) *interpolation*, *transposed multi-point evaluation*, and *transposed interpolation*. For fixed $\alpha_1, \ldots, \alpha_n$, these complexities can often be further reduced to $O(\mathsf{M}(n) \log n / \log \log n)$ using techniques from [45].

For our factorization algorithms, we will sometimes need to assume the existence of an algorithm to factor univariate polynomials in $\mathbb{K}[x]$ into irreducible factors. We will denote by $\mathsf{F}(d)$ the cost of such a factorization as a function of degree $d$. We will always assume that $\mathsf{F}(d)/d$ is non-decreasing. In particular $\mathsf{F}(d_1) + \mathsf{F}(d_2) \leqslant \mathsf{F}(d_1 + d_2)$ for all $d_1$ and $d_2$.

If $\mathbb{K}$ is the finite field $\mathbb{F}_q$ with $q = p^{\kappa}$ elements for some odd $q$, then the best univariate factorization methods are randomized of Las Vegas type. When allowing for such algorithms, we may take $\mathsf{F}(d) = O(d \,\mathsf{M}(d) \log (qd))$, by using Cantor and Zassenhaus' method from [16]. With the use of fast modular composition [70], we may take

$$\mathsf{F}(d) = d^{1.5+o(1)} \log^{1+o(1)} q + \tilde{O}(d \log^2 q),$$

but this algorithm is only relevant in theory, for the moment [50]. If the extension degree $\kappa$ is composite, then this can be exploited to lower the practical complexity of factorization [53].

## 2.2. The Schwarz–Zippel lemma

In all probabilistic algorithms in this paper, $N$ will stand for a sufficiently large integer such that "random elements in $\mathbb{K}^{\neq}$" are chosen in some fixed subset of $\mathbb{K}^{\neq}$ of size at least $N$. In the case when $N$ is larger than the cardinality $|\mathbb{K}^{\neq}|$ of $\mathbb{K}^{\neq}$, this means that $\mathbb{K}$ needs to be replaced by an algebraic extension of degree $>\log N / \log |\mathbb{K}|$, which induces a logarithmic $\tilde{O}(\log N)$ overhead for the cost of field operations in $\mathbb{K}$. We will frequently use the following well-known lemma:

LEMMA 2.1. (SCHWARZ [99]–ZIPPEL [108]) *Let $P \in \mathbb{K}[x_1, \ldots, x_n]$ be a polynomial of total degree $d$. Let $S \subseteq \mathbb{K}$ be finite and let $\alpha_1, \ldots, \alpha_n \in S$ be chosen independently and uniformly. Then the probability that $P(\alpha_1, \ldots, \alpha_n) = 0$ is at most $d / |S|$.* □

COROLLARY 2.2. *Let $s \in \mathbb{N}^{>}$. The probability that $P(\alpha_1^k, \ldots, \alpha_n^k) = 0$ for some $k \in \{1, \ldots, s\}$ is at most $d \binom{s}{2} / |S|$.*

**Proof.** We apply the lemma to $P(x_1, \ldots, x_n)\, P(x_1^2, \ldots, x_n^2) \cdots P(x_1^s, \ldots, x_n^s)$. □

COROLLARY 2.3. *Let $s \in \mathbb{N}^{>}$. Let $\alpha_1, \ldots, \alpha_n \in \mathbb{K}^{\neq}$ and let $\beta_1, \ldots, \beta_n \in \mathbb{K}^{\neq}$ be chosen independently at random. Then the probability that $P(\alpha_1^k \beta_1, \ldots, \alpha_n^k \beta_n) = 0$ for some $k \in \{1, \ldots, s\}$ is at most $d s / |S|$.*

**Proof.** We apply the lemma to $P(\alpha_1 x_1, \ldots, \alpha_n x_n)\, P(\alpha_1^2 x_1, \ldots, \alpha_n^2 x_n) \cdots P(\alpha_1^s x_1, \ldots, \alpha_n^s x_n)$. □

## 2.3. Sparse polynomial interpolation

Consider a polynomial $F \in \mathbb{K}[x_1, \ldots, x_n]$ that is presented as a blackbox function. The task of *sparse interpolation* is to recover the sparse representation (1.1) from a sufficient number of blackbox evaluations of $F$. One may distinguish between the cases when the exponents $\gamma_1, \ldots, \gamma_s$ of $F$ are already known or not. (Here "known exponents" may be taken liberally to mean a superset of reasonable size of the set of actual exponents.)

One popular approach for sparse interpolation is based on Prony's geometric sequence technique [91, 5]. This approach requires an *admissible ratio* $\alpha = (\alpha_1, \ldots, \alpha_n) \in (\mathbb{K}^{\neq})^n$, such that for any $k_1, \ldots, k_n \in \mathbb{N}$, there is an algorithm to recover $k_1, \ldots, k_n$ from $\alpha_1^{k_1} \cdots \alpha_n^{k_n}$. If $\operatorname{char} \mathbb{K} = 0$, then we may take $\alpha_i$ to be the $i$-th prime number, and use prime factorization in order to recover $k_1, \ldots, k_n$. If $\mathbb{K} = \mathbb{F}_p$ is a finite field, where $p$ is a smooth prime (i.e. $p - 1$ has many small prime divisors), then one may recover exponents using the tangent Graeffe method [38].

Given an admissible ratio $\alpha$, Prony's method allows for the sparse interpolation of $F$ using $2s$ evaluations of $F$ at $\alpha^i := (\alpha_1^i, \ldots, \alpha_n^i)$ for $i = 0, \ldots, 2s - 1$, as well as $O(\mathsf{M}(s) \log s)$ operations in $\mathbb{K}$ for determining $\alpha^{\gamma_1}, \ldots, \alpha^{\gamma_s}$, and $s$ subsequent exponent recoveries. If $\mathbb{K} = \mathbb{F}_q$, then the exponents can be recovered if $q > \max(2s, d)$, which can be ensured by working over a field extension $\mathbb{F}_{q^\kappa}$ of $\mathbb{F}$ with $\kappa = O(\log s + \log d)$. If the exponents $\gamma_1, \ldots, \gamma_s$ are already known, then the coefficients can be obtained from $s$ evaluations of $F$ at $\alpha^0, \ldots, \alpha^{s-1}$ using one transposed univariate interpolation of cost $O(\mathsf{M}(s) \log s)$.

If $\mathbb{K}$ is a finite field, then it can be more efficient to consider an *FFT ratio* $\alpha$ for which $\alpha_1, \ldots, \alpha_n$ are roots of unity. When choosing these roots of unity with care, possibly in an extension field of $\mathbb{K}$, sparse interpolation can often be done in time $O(\mathsf{M}(s))$ from $O(s)$ values of $F$, using discrete Fourier transforms; see [52, 46] for details.

In what follows, we will denote by $\mathsf{S}(s)$ the cost of sparse interpolation of size $s$, given $O(s)$ values of $F$ at a suitable geometric sequence. When using Prony's approach, we may thus take $\mathsf{S}(s) = O(\mathsf{M}(s) \log s)$, whenever the cost to recover the exponents $k_1, \ldots, k_n$ from $\alpha_1^{k_1} \cdots \alpha_n^{k_n}$ is negligible. If the discrete Fourier approach is applicable, then we may even take $\mathsf{S}(s) = O(\mathsf{M}(s))$.

**Remark 2.4.** It would be more accurate to write $\mathsf{S}(s, n, d)$ for the cost of sparse interpolation, since this cost may also depend on the dimension $n$ and the total degree $d$. In particular, manipulations of the exponent vectors typically add $\Omega(s n \log d)$ to the bit complexity. Nonetheless, in our algebraic complexity model, this cost can often be discarded: if char $\mathbb{K} = 0$, then the sparse interpolation of the exponents only gives rise to a constant overhead using the derivative trick from [57]. In practice, this remains true as long as the exponent vectors can be packed into small integers, such as 64 bit machine integers. In order to enhance the readability of our complexity analyses, we therefore assume that we are in a regime where the cost $\mathsf{S}(s)$ of sparse interpolation can be modeled in terms of the sole parameter $s$. For recent theoretical work on sparse interpolation when $n$ and/or $d$ are allowed to become large, we refer to [36, 54].

**Remark 2.5.** If we have a bound $s \geqslant s_F$ for the number of terms of $F$, then we assume that our sparse interpolation method is deterministic and that it interpolates $F$ in time $\mathsf{S}(s)$. If we do not know the number of terms of $F$, then we may run the interpolation method for a guessed number of $s$ terms. We may check the correctness of our guessed interpolation $\tilde{F}$ by verifying that the evaluations of $F$ and $\tilde{F}$ coincide at a randomly chosen point. By the Schwartz-Zippel lemma, this strategy succeeds with probability at least $1 - d/N$.

**Remark 2.6.** The above interpolation methods readily generalize to the case when we use a geometric progression of the form $\beta, \beta\alpha, \beta\alpha^2, \ldots$ with $\beta \in (\mathbb{K}^{\neq})^n$ as the evaluation points, by considering the function $g(x_1, \ldots, x_n) = f(\beta_1 x_1, \ldots, \beta_n x_n)$ instead of $f$. Taking a random $\beta$ avoids certain problems due to degeneracies; this is sometimes called "diversification" [35]. If $\alpha$ is itself chosen at random, then it often suffices to simply take $\beta = \alpha$. We will occasionally do so without further mention.

**Remark 2.7.** For many probabilistic proofs in the sequel of this paper, we will rely on Corollary 2.2. However, this requires the ratios $\alpha$ of geometric progressions to be picked at random, which excludes FFT ratios. Alternatively, we could have relied on Corollary 2.3 and diversification of all input and output polynomials for a fixed random scaling factor $\beta \in (\mathbb{K}^{\neq})^n$ (see the above remark).

Assume for instance that we wish to factor $F \in \mathbb{K}[x_1, \ldots, x_n]$. Then the factors $A \in \mathbb{K}[x_1, \ldots, x_n]$ of $F$ are in one-to-one correspondence with the factors $A(\beta_1 x_1, \ldots, \beta_n x_n)$ of $F(\beta_1 x_1, \ldots, \beta_n x_n)$. If we rely on Corollary 2.3 instead of 2.2 in our complexity bounds for factoring $F$ (like the bounds in Theorems 6.11 or 7.2 below), then the cost of diversification adds an extra term $O(\bar{s} d)$, where $d := \deg F$ and $\bar{s}$ is the total size of $F$ and the computed factors. On the positive side, in the corresponding bounds for the probability of failure, the quadratic dependence $\binom{s}{2}$ on the number $s$ of evaluation points reduces to a linear one. Similar adjustments apply for other operations such as gcd computations.

## 2.4. Sparse evaluation at geometric progressions

Opposite to sparse interpolation, one may also consider the evaluation of $F$ at $s$ points $\alpha^0, \ldots, \alpha^{s-1}$ in a geometric progression. In general, this can be done in time $O(\mathsf{M}(s) \log s)$, using one transposed multi-point evaluation of size $s$. If $\alpha$ is a suitable FFT ratio, then this complexity drops to $O(\mathsf{M}(s))$, using a discrete Fourier transform of size $O(s)$. In what follows, we will assume that this operation can always be done in time $\mathsf{S}(s)$.

**Remark 2.8.** Strictly speaking, the cost of multipoint evaluation on a geometric sequence may also depend on $n$ and $d$, so similar observations as in Remark 2.4 apply.

More generally, we may consider the evaluation of $F$ at $t$ points $\alpha^0, \ldots, \alpha^{t-1}$ in a geometric progression. If $t > s$, we may do this in time $\mathsf{S}(s)\, t / s + O(t)$, by reducing to the evaluation of $F$ at $\lceil t / s \rceil$ progressions of size $s$. To obtain the evaluations of $F$ at $\alpha^i, \ldots, \alpha^{i+s-1}$ for $i > 0$, we can evaluate $F \circ (\alpha^i x)$ at $\alpha^0, \alpha^1, \ldots, \alpha^{s-1}$. If $s > t$, then we may cut $F$ into $\lceil s / t \rceil$ polynomials of size $\leqslant t$, and do the evaluation in time $\mathsf{S}(t)\, s / t + O(s)$.

**Remark 2.9.** If $\alpha$ is an FFT-ratio, then the bound for the case when $s > t$ further reduces to $O(s + \mathsf{S}(t))$ plus $O(d s)$ bit operations on exponents, since the cyclic projections from [46, 52] reduce $F$ in linear time to cyclic polynomials with $O(t)$ terms before applying the FFTs. We did not exploit this optimization for the complexity bounds in this paper, since we preferred to state these bounds for general ratios $\alpha$, but it should be straightforward to adapt them to the specific FFT case.

In this paper, we will frequently need to evaluate $F$ at all but one or two variables. Assume for instance that

$$F(x_1, \ldots, x_n) \;=\; F_0(x_2, \ldots, x_n) + F_1(x_2, \ldots, x_n)\, x_1 + \cdots + F_\delta(x_2, \ldots, x_n)\, x_1^\delta,$$

where $F_k$ has $s_k$ terms for $k = 0, \ldots, \delta$ and $s := s_0 + \cdots + s_\delta$. Then using the above method, we can compute $F(x_1, \alpha_2^i, \ldots, \alpha_n^i)$ for $i = 0, \ldots, t-1$ using $\mathsf{S}(t)\, (\lceil s_0 / t \rceil + \cdots + \lceil s_\delta / t \rceil) + O(s) \leqslant \mathsf{S}(t)\, (s / t + \delta + 1) + O(s)$ operations.

One traditional application of the combination of sparse evaluation and interpolation are probabilistic algorithms for multiplication and exact division of sparse multivariate polynomials. For the given $A, B \in \mathbb{K}[x_1, \ldots, x_n]$, we can compute the product $C = A B$ by evaluating $A, B$, and $C$ at a geometric progression $\alpha^0, \ldots, \alpha^{m-1}$ with $m = O(s_C)$ and recovering of $C$ in the sparse representation (1.1) via sparse interpolation. The total cost of this method is bounded by $O(\mathsf{S}(\bar{s}))$ operations in $\mathbb{K}$, where $\bar{s} := \max(s_A, s_B, s_C)$. If $C$ and $A$ are known, then the exact quotient $B = C / A$ can be computed in a similar fashion and with the same complexity. If $s := s_B \ll \bar{s}$, then the quotient $B$ can actually be computed in time $O(\bar{s}\, \mathsf{M}(s) / s)$. Divisions by zero are avoided through diversification, with overhead $O(n \bar{s})$ and probability at least $1 - d_A \binom{s_B}{2} / N$, by Corollary 2.2.

## 2.5. Newton polytopes

Consider a multivariate polynomial $P \in \mathbb{K}[x_1, \ldots, x_n]$. We define $\operatorname{hull} P \subseteq \mathbb{R}^n$ to be the convex hull of $\operatorname{supp} P$ and we call it the *Newton polytope* of $P$. Given another polynomial $Q \in \mathbb{K}[x_1, \ldots, x_n]$, it is well known that

$$\operatorname{hull} P Q \;=\; \operatorname{hull} P + \operatorname{hull} Q,$$

where the *Minkowski sum* of two subsets $S, T \subseteq \mathbb{R}^n$ is $S + T := \{s + t : s \in S, t \in T\}$.

Let $w = (w_1, \ldots, w_n) \in (\mathbb{Z}^n)^{\neq}$ be a non-zero weight vector. We define the *w-degree*, *w-valuation*, and *w-ecart* of a non-zero polynomial $P \in \mathbb{K}[x_1, \ldots, x_n]$ by

$$\begin{aligned}
\deg_w P &:= \max_{(e_1, \ldots, e_n) \in \operatorname{supp} F} (w_1 e_1 + \cdots + w_n e_n) \\
\operatorname{val}_w P &:= \min_{(e_1, \ldots, e_n) \in \operatorname{supp} F} (w_1 e_1 + \cdots + w_n e_n) \\
\operatorname{ec}_w P &:= \deg_w P - \operatorname{val}_w P.
\end{aligned}$$

Note that $\operatorname{val}_w P := -\deg_{-w} P$ and $\operatorname{ec}_w P = \operatorname{ec}_{-w} P$, where we exploit the fact that we allow for negative weights. We say that $P$ is *w-homogeneous* if $\deg_w P = \operatorname{val}_w P$. Any $P$ can uniquely be written as a sum

$$P \;=\; P_{\deg_w P; w} + \cdots + P_{\operatorname{val}_w P; w},$$

of its *w-homogeneous parts*

$$P_{i;w} := \sum_{e \in \operatorname{supp} P, w_1 e_1 + \cdots + w_n e_n = i} P_e x^e.$$

We call $\operatorname{lp}_w P := P_{\deg_w P; w}$ and $\operatorname{tp}_w P := P_{\operatorname{val}_w P; w}$ the *w-leading* and *w-trailing* parts of $P$. We say that $P$ is *w-regular* if $\operatorname{lp}_w P$ consists of a single term $c x^i$. In that case, we denote $\operatorname{lc}_w P := c$ and we say that $P$ is *w-monic* if $c = 1$. Given another non-zero polynomial $Q \in \mathbb{K}[x_1, \ldots, x_n]$, we have

$$\begin{aligned} \operatorname{lp}_w PQ &= (\operatorname{lp}_w P)(\operatorname{lp}_w Q) \\ \operatorname{tp}_w PQ &= (\operatorname{tp}_w P)(\operatorname{tp}_w Q) \end{aligned}$$

Setting $\mathcal{H}_{\lambda;w} := \{e \in \mathbb{R}^n : e_1 w_1 + \cdots + e_n w_n = \lambda\}$, we also have

$$\begin{aligned} \operatorname{hull} \operatorname{lp}_w P &= \operatorname{hull} P \cap \mathcal{H}_{\deg_w P; w} \\ \operatorname{hull} \operatorname{tp}_w P &= \operatorname{hull} P \cap \mathcal{H}_{\operatorname{val}_w P; w}. \end{aligned}$$

The Newton polytopes $\operatorname{hull} \operatorname{lp}_w P$ and $\operatorname{hull} \operatorname{tp}_w P$ are *facets* of $\operatorname{hull} P$. In particular, they are contained in the boundary $\partial \operatorname{hull} P$.

## 2.6. Laurent polynomials

Consider the rings $\mathbb{P} := \mathbb{P}_n := \mathbb{K}[x_1, \ldots, x_n]$ and $\mathbb{L} := \mathbb{L}_n := \mathbb{K}[x_1, x_1^{-1}, \ldots, x_n, x_n^{-1}] = \mathbb{P}_n x_1^{\mathbb{Z}} \cdots x_n^{\mathbb{Z}}$ of ordinary polynomials and *Laurent polynomials*. Both rings are unique factorization domains, but the group of units of $\mathbb{P}$ is $\mathbb{K}^{\neq}$, whereas the group of units of $\mathbb{L}$ is $\mathbb{K}^{\neq} x_1^{\mathbb{Z}} \cdots x_n^{\mathbb{Z}}$. Factoring in $\mathbb{P}$ is therefore essentially the same as factoring in $\mathbb{L}$ up to multiplications with monomials in $x_1^{\mathbb{Z}} \cdots x_n^{\mathbb{Z}}$. For instance, the factorization $5 \cdot x_1 \cdot x_1 \cdot x_2 \cdot (x_1 - x_2) \cdot (7 x_1 + x_2^2 - x_3^3)$ in $\mathbb{P}$ gives rise to the factorization $(5 x_1^2 x_2) \cdot (x_1 - x_2) \cdot (7 x_1 + x_2^2 - x_3^3)$ in $\mathbb{L}$. Conversely, the factorization $(5 x_1 x_2^2) \cdot (x_1 x_2^{-1} - 1) \cdot (7 x_1^2 + x_1 x_2^2 - x_1 x_3^3)$ in $\mathbb{L}$ gives rise to the factorization $5 \cdot x_1 \cdot x_1 \cdot x_2 \cdot (x_1 - x_2) \cdot (7 x_1 + x_2^2 - x_3^3)$ in $\mathbb{P}$. Similarly, computing gcds in $\mathbb{P}$ is essentially the same problem as computing gcds in $\mathbb{L}$.

Given any $m \times n$ matrix $M \in \mathbb{Z}^{m \times n}$, we define the *monomial map* $\varphi_M : \mathbb{L}_n \to \mathbb{L}_m$ by

$$\varphi_M(P(x_1, \ldots, x_n)) = P(x_1^{M_{1,1}} \cdots x_m^{M_{m,1}}, \ldots, x_1^{M_{1,n}} \cdots x_m^{M_{m,n}}).$$

This is clearly a homomorphism, which is injective (or surjective) if the linear map $\mathbb{Z}^n \to \mathbb{Z}^m; a \mapsto M a$ is injective (or surjective). Note also that $\varphi_{MN} = \varphi_M \circ \varphi_N$ for any matrices $M \in \mathbb{Z}^{m \times n}$ and $N = \mathbb{Z}^{n \times r}$. In particular, if $M \in \mathbb{Z}^{n \times n}$ is unimodular, then $\varphi_M$ is an automorphism of $\mathbb{L}_n$ with $\varphi_M^{-1} = \varphi_{M^{-1}}$.

Laurent polynomials are only slightly more general than ordinary polynomials and we already noted above that factoring in $\mathbb{P}_n$ is essentially the same problem as factoring in $\mathbb{L}_n$ (and similarly for gcd computations). It is also straightforward to adapt the definitions from Section 2.5 and most algorithms for sparse interpolation to this slightly more general setting. The main advantage of Laurent polynomials is that they are closed under monomial maps, which allows us to change the geometry of the support of a polynomial without changing its properties with respect to factorization.

## 2.7. Tagging

Let $w \in (\mathbb{Z}^n)^{\neq}$ be a non-zero weight vector and let $\mathbb{L}_n^{\#} := \mathbb{K}[x_1, x_1^{-1}, \ldots, x_n, x_n^{-1}, t, t^{-1}]$. We define the *w-tagging map* by

$$\begin{aligned} \tau_w : \mathbb{L}_n &\longrightarrow \mathbb{L}_n^{\#} \\ P(x_1, \ldots, x_n) &\longmapsto P(x_1 t^{w_1}, \ldots, x_n t^{w_n}). \end{aligned}$$

This map is an injective monomial map. For any $P \in \mathbb{K}[x_1, \ldots, x_n]$, we have $\deg_t \tau_w(P) = \deg_w P$, $\mathrm{val}_t \tau_w(P) = \mathrm{val}_w P$, $s_{\tau_w(P)} = s_P$, and $\mathrm{ec}_t \tau_w(P) := \deg_t \tau_w(P) - \mathrm{val}_t \tau_w(P) = \mathrm{ec}_w P$. Divisibility and gcds are preserved as follows:

LEMMA 2.10. *Let $P, Q, G \in \mathbb{P}_n$ with $\mathrm{val}_{x_i} P = \mathrm{val}_{x_i} Q = \mathrm{val}_{x_i} G = 0$ for $i = 1, \ldots, n$. Then*

*a) $P$ divides $Q$ in $\mathbb{P}_n$ if and only if $\tau_w(P)$ divides $\tau_w(Q)$ in $\mathbb{L}_n^\#$.*

*b) $G = \gcd(P, Q)$ in $\mathbb{P}_n$ if and only if $\tau_w(G) = \gcd(\tau_w(P), \tau_w(Q))$ in $\mathbb{L}_n^\#$.*

**Proof.** We claim that $P$ divides $Q$ in $\mathbb{P}_n$ if and only if $P$ divides $Q$ in $\mathbb{L}_n$. One direction is clear. Assume that $P$ divides $Q$ in $\mathbb{L}_n$, so that $Q = A P$ with $A \in \mathbb{L}_n$. We may uniquely write $A = x^e A'$ with $e \in \mathbb{Z}^n$ and $A' \in \mathbb{P}_n$ such that $\mathrm{val}_{x_i} A' = 0$ for $i = 1, \ldots, n$. Since $0 = \mathrm{val}_{x_i} Q = \mathrm{val}_{x_i} A + \mathrm{val}_{x_i} P = \mathrm{val}_{x_i} = e_i + \mathrm{val}_{x_i} A' = e_i$ for $i = 1, \ldots, n$, it follows that $e = 0$. Hence $P$ divides $Q$ in $\mathbb{P}_n$. Our claim implies that $P$ divides $Q$ in $\mathbb{P}_n$ if and only if $P$ divides $Q$ in $\mathbb{L}_n^\#$. Now we may further extend $\tau_w$ to a monomial automorphism of $\mathbb{L}_n^\#$ by sending $t$ to itself. This yields $(a)$. The second property is an easy consequence. □

COROLLARY 2.11. *Let $P, Q \in \mathbb{P}_n$ and $G = \gcd(P, Q)$. Let $\hat{G} = \gcd(\hat{P}, \hat{Q})$, where $\hat{P} := \tau_w(P)$ and $\hat{Q} := \tau_w(Q)$. Let $\nu \in \mathbb{Z}^n$ be such that $\nu_i := \min(\mathrm{val}_{x_i} P, \mathrm{val}_{x_i} Q) - \mathrm{val}_{x_i} \hat{G}$ for $i = 1, \ldots, n$. Then*

$$G(x_1, \ldots, x_n) = x^\nu \hat{G}(x_1, \ldots, x_n, 1). \qquad \square$$

Given a $w$-regular polynomial $F \in \mathbb{K}[x_1, \ldots, x_n]$, we note that any divisor $P | F$ must again be $w$-regular. Hence, modulo multiplication with a suitable constant in $\mathbb{K}$, we may always normalize such a divisor to become $w$-monic. In the setting of Laurent polynomials, we may further multiply $P$ by a monomial in $x_1^{\mathbb{Z}} \cdots z_n^{\mathbb{Z}}$ such that $\mathrm{lp}_w P = 1$. Similarly, when applying $\tau_w$, we can always normalize $\tau_w(P)$ to be monic as a Laurent polynomial in $t$ by considering $(\mathrm{lp}_w P)^{-1} \tau_w(P)$.

Both for gcd computations and factorization, this raises the question of how to find weights $w$ for which a given non-zero polynomial $F \in \mathbb{K}[x_1, \ldots, x_n]$ is $w$-regular. In [51, Section 4.3], a way was described to compute such a *regularizing weight* $w$: let $i = (i_1, \ldots, i_n) \in \mathrm{supp}\, F$ be such that $i_1^2 + \cdots + i_n^2$ is maximal. In that case, it suffices to take $w := i$, and we note that $\mathrm{ec}_w F \leqslant d^2$, where $d$ is the total degree of $F$. For our applications in Sections 3.2 and 5 it is often important to find a $w$ for which $\mathrm{ec}_w F$ is small. By trying a few random weights with small (possibly negative) entries, it is often possible to find a regularizing weight $w$ with $\mathrm{ec}_w P = O(1)$ or $\mathrm{ec}_w P = O(d)$.

**Example 2.12.** Consider $F = 2 x^2 y + 3 x y^2 + x y + 3 y + 2 z + 4 \in \mathbb{Q}[x, y, z]$. Then, $F$ is not $w_1$-regular for the natural weight $w_1 := (1, 1, 1)$. If we take $w_2 := (2, 1, 0)$ instead, then $F$ is $w_2$-regular with $\mathrm{ec}_{w_2} F = 5$, and we arrive at

$$\tau_{w_2}(F) = (2 x^2 y) t^5 + (3 x y^2) t^4 + (x y) t^3 + (3 y) t + 2 z + 4.$$

Furthermore, $\tau_{w_2}(F)$ can be normalized to be monic as a Laurent polynomial in $t$ by considering $\tau_{w_2}(F) / (2 x^2 y)$. Note that $w_3 := (0, 0, 1)$ is also a regularizing weight for $F$ with $\mathrm{ec}_{w_3} F = 1$.

## 3. MULTIVARIATE GCD COMPUTATIONS

Before studying the factorization problem for multivariate polynomials, it is interesting to consider the easier problem of gcd computations. In this section we introduce two approaches for gcd computations that will also be useful later for factoring polynomials.

The first approach is iterative on the number of variables. It will be adapted to the factorization problem in Section 6. The second approach is more direct, but requires a regularizing weight (see Section 2.7). Square-free factorization can be accomplished using a similar technique, as we shall see in Section 5. The algorithms from Section 7 also draw some of their inspiration from this technique, but also rely on Newton polygons instead of regularizing weights.

## 3.1. Iterative computation of gcds

Let $c_1,\ldots,c_n$ be random elements of $\mathbb{K}^{\neq}$. For any $A\in\mathbb{K}[x_1,\ldots,x_n]$ and $k=1,\ldots,n$, we define

$$A^{[k]}(x_1,\ldots,x_k) \;:=\; A(x_1,\ldots,x_k,c_{k+1},\ldots,c_n).$$

Let $P,Q\in\mathbb{K}[x_1,\ldots,x_k]^{\neq}$and $G:=\gcd(P,Q)$. As we will see below, $G^{[k]}=\gcd(P^{[k]},Q^{[k]})$ for $k=1,\ldots,n$ with high probability. We may easily compute the univariate greatest common divisor $G^{[1]}:=\gcd(P^{[1]},Q^{[1]})$. In this subsection, we shall describe an iterative algorithm to compute $G^{[k+1]}$ from $G^{[k]}$ for $k=1,\ldots,n-1$. Eventually, this yields $G=G^{[n]}$.

Let $\alpha=(\alpha_1,\ldots,\alpha_k)$ be an admissible ratio or an FFT ratio. For any $A\in\mathbb{K}[x_1,\ldots,x_n]$ and any $i\in\mathbb{N}$, let

$$\begin{aligned} A^{\langle k+1,i\rangle}(u) &:= A^{[k+1]}(\alpha_1^i,\ldots,\alpha_k^i,u)\\ A^{[k,i]} &:= A^{[k]}(\alpha_1^i,\ldots,\alpha_k^i). \end{aligned}$$

For any $i\in\mathbb{N}$, we have $G^{\langle k+1,i\rangle}=\gcd(P^{\langle k+1,i\rangle},Q^{\langle k+1,i\rangle})$ with high probability. Now these greatest common divisors are only defined up to non-zero scalar multiples in $\mathbb{K}^{\neq}$. Nevertheless, there exists a unique greatest common divisor $G^{[k+1]}$ of $P^{[k+1]}$ and $G^{[k+1]}$ whose evaluation at $x_{k+1}:=c_{k+1}$ coincides with $G^{[k]}$.

If $G^{[k]}$ is known, then $G^{[k,i]}$, $P^{\langle k+1,i\rangle}$, and $Q^{\langle k+1,i\rangle}$ can be computed for successive $i\in\mathbb{N}$ using fast evaluation at geometric progressions. For any $i\in\mathbb{N}$, we may then compute the univariate gcd $G^{\langle k+1,i\rangle}$ of $P^{\langle k+1,i\rangle}$ and $Q^{\langle k+1,i\rangle}$, under the normalization constraint that $G^{\langle k+1,i\rangle}(c_{k+1})=G^{[k,i]}$. It finally suffices to interpolate $G^{[k+1]}$ from sufficiently many $G^{\langle k+1,i\rangle}$. This yields the following algorithm:

**Algorithm 3.1**
**Input:** $P,Q\in\mathbb{K}[x_1,\ldots,x_n]^{\neq}$
**Output:** $G\in\mathbb{K}[x_1,\ldots,x_n]$, such that $G=\gcd(P,Q)$

If $n\leqslant 1$, then compute $\gcd(P,Q)$ using a univariate algorithm and return it
Compute $G^{[n-1]}$ by recursively applying the algorithm to $P^{[n-1]}$ and $Q^{[n-1]}$
Let $m:=s_{G^{[n-1]}}$
Compute $G^{[n-1,i]}$, $P^{\langle n,i\rangle}$, $Q^{\langle n,i\rangle}$ for $i=1,\ldots,m$ using sparse evaluation
Compute $G^{\langle n,i\rangle}=\gcd(P^{\langle n,i\rangle},Q^{\langle n,i\rangle})$ with $G^{\langle n,i\rangle}(c_n)=G^{[n-1,i]}$ for $i=1,\ldots,m$
Recover $G$ from $G^{\langle n,1\rangle},\ldots,G^{\langle n,m\rangle}$ using sparse interpolation
Return $G$

Before we analyze this algorithm, we will need a few probabilistic lemmas. Assume that $c_1,\ldots,c_n$ and $\alpha_1,\ldots,\alpha_n$ are independently chosen at random from a subset of $\mathbb{K}^{\neq}$ of size at least $N$ (if $\mathbb{K}$ is a small finite field, this forces us to move to a field extension).

LEMMA 3.1. *Let $A,B\in\mathbb{K}[x_1,\ldots,x_n]$ be such that $A^{[k+1]}$ and $B^{[k+1]}$ are coprime. Then the probability that $A^{[k]}$ and $B^{[k]}$ are not coprime is bounded by $2k\,d_A\,d_B/N$.*

**Proof.** Let $x_i$ with $i \leqslant k$ be a variable that occurs both in $A^{[k+1]}$ and in $B^{[k+1]}$. Then

$$R_i \ := \ \operatorname{Res}_{x_i}(A^{[k+1]}, B^{[k+1]}) \ \neq \ 0.$$

If $x_i$ occurs in $\gcd(A^{[k]}, B^{[k]})$, then $R_i(x_1, \ldots, x_k, c_{k+1}) = 0$, which can happen for at most $\deg R_i \leqslant \deg_{x_i} A^{[k+1]} \deg B^{[k+1]} + \deg A^{[k+1]} \deg_{x_i} B^{[k+1]} \leqslant 2 d_A d_B$ values of $c_{k+1}$. □

LEMMA 3.2. *Let $P, Q \in \mathbb{K}[x_1, \ldots, x_n]$ and $G = \gcd(P, Q)$. Then the probability that $G^{[k]} \neq \gcd(P^{[k]}, Q^{[k]})$ for some $k \in \{1, \ldots, n\}$ is bounded by $n^2 d_P d_Q / N$.*

**Proof.** Let us write $P = A G$ and $Q = B G$, where $A$ and $B$ are coprime. Then we have $G^{[k]} \neq \gcd(P^{[k]}, Q^{[k]})$ if and only if $A^{[k]}$ and $B^{[k]}$ are not coprime. The result now follows by applying the previous lemma for $k = n-1, \ldots, 1$. □

THEOREM 3.3. *Let $s := s_G$, $\bar{s} := s_P + s_Q + s_G$, $d := \max(d_P, d_Q)$, and $\delta := \max(\delta_P, \delta_Q)$. Then Algorithm 3.1 is correct with probability at least $1 - \left(n^2 d^2 + n \binom{s}{2} \delta\right) / N$ and it runs in time*

$$O(n\,((\bar{s}/s + \delta)\,\mathsf{S}(s) + s\,\mathsf{M}(\delta) \log \delta)).$$

**Remark.** The probability bound implicitly assumes that $N > n^2 d^2 + n \binom{s}{2} \delta$, since the statement becomes void for smaller $N$. In particular, we recall that this means that the cardinality of $\mathbb{K}$ should be at least $n^2 d^2 + n \binom{s}{2} \delta$.

**Proof.** Assuming that $G^{[k]} = \gcd(P^{[k]}, Q^{[k]})$ and $G^{[k,i]} \neq 0$ for $k = 1, \ldots, n-1$ and $i = 1, \ldots, s$, let us prove that Algorithm 3.1 returns the correct answer. Indeed, these assumptions imply that $\Gamma := \gcd(P^{\langle n,i \rangle}, Q^{\langle n,i \rangle})$ can be normalized such that $\Gamma(c_n) = G^{[n-1,i]}$ and that the unique such $\Gamma$ must coincide with $G^{\langle n,i \rangle}$. Now $G^{[k]} = \gcd(P^{[k]}, Q^{[k]})$ fails for some $k$ with probability at most $n^2 d^2 / N$, by Lemma 3.2. Since $\delta_G \leqslant \delta$, the condition $G^{[k,i]} \neq 0$ fails with probability at most $n \binom{s}{2} \delta / N$ for some $k$ and $i$, by Corollary 2.2. This completes the probabilistic correctness proof.

As to the complexity, let us first ignore the cost of the recursive call. Then the sparse evaluations of $G^{[n-1,i]}$, $P^{\langle n,i \rangle}$, and $Q^{\langle n,i \rangle}$ can be done in time $O((\bar{s}/s + \delta)\,\mathsf{S}(s))$: see Section 2.4. The univariate gcd computations take $O(s\,\mathsf{M}(\delta) \log \delta)$ operations. Finally, the recovery of $G$ using sparse interpolation can be done in time $O(\delta\,\mathsf{S}(s))$. Altogether, the complexity without the recursive calls is bounded by $O((\bar{s}/s + \delta)\,\mathsf{S}(s) + s\,\mathsf{M}(\delta) \log \delta)$. We conclude by observing that the degrees and numbers of terms of $P$, $Q$, and $G$ can only decrease during recursive calls. Since the recursive depth is $n$, the complexity bound follows. □

**Remark 3.4.** When recovering $G$ from $G^{\langle n,1 \rangle}, \ldots, G^{\langle n,m \rangle}$ using sparse interpolation, one may exploit the fact that the exponents of $x_1, \ldots, x_{n-1}$ in $G$ are already known.

**Example 3.5.** Let $A, B, C \in \mathbb{K}[x_1, \ldots, x_n]$ be random polynomials of total degree $d$ and consider $P := A B$, $Q := A C$. With high probability, $G := \gcd(P, Q) = A$. Let us measure the overhead of recursive calls in Algorithm 3.1 with respect to the number of variables $n$. With high probability, we have

$$s_G \ = \ \binom{n+d}{n}, \qquad s_{G^{[n-1]}} \ = \ \binom{n-1+d}{n-1} \ = \ \frac{n+d}{n}\, s_{G^{[n-1]}}$$

and

$$s_P = s_Q = \binom{n+2d}{n}, \qquad s_P = \frac{n+2d}{n} s_{P^{[n-1]}}, \qquad s_Q = \frac{n+2d}{n} s_{Q^{[n-1]}}.$$

Assuming that $d \geqslant n$, it follows that

$$2 s_{G^{[n-1]}} \leqslant s_G, \qquad 3 s_{P^{[n-1]}} \leqslant s_P, \qquad 3 s_{Q^{[n-1]}} \leqslant s_Q.$$

This shows that the sizes of the supports of the input and output polynomials in Algorithm 3.1 become at least twice as small at every recursive call. Consequently, the overall cost is at most twice the cost of the top-level call, roughly speaking.

## 3.2. Gcd computations through regularizing weights

Algorithm 3.1 has the disadvantage that the complexity bound in Theorem 3.3 involves a factor $n$. Let us now present an alternative algorithm that avoids this pitfall, but which may require a non-trivial monomial change of variables. Our method is a variant of the algorithm from [51, Section 4.3].

Given $P, Q \in \mathbb{K}[x_1, \ldots, x_n]$, we first compute a regularizing weight $w$ for $P$ or $Q$, for which $e := \max(\mathrm{ec}_w P, \mathrm{ec}_w Q)$ is as small as possible. From a practical point of view, as explained in Section 2.7, we first try a few random small weights $w$. If no regularizing weight is found in this way, then we may always revert to the following choice:

LEMMA 3.6. *For vectors $\nu \in \mathbb{R}^n$, let $|\nu| := \sqrt{\nu_1^2 + \cdots + \nu_n^2}$. Let $i = (i_1, \ldots, i_n) \in \operatorname{supp} P$ be such that $|i| \leqslant d_P$ is maximal. Let $j = (j_1, \ldots, j_n) \in \operatorname{supp} Q$ be such that $|j| \leqslant d_Q$ is maximal. Let $w := i$ if $|i| \leqslant |j|$ and $w := j$ otherwise. Then $e := \max(\mathrm{ec}_w P, \mathrm{ec}_w Q) \leqslant d_P d_Q$.*

**Proof.** Assume that $w = i$. Then $w \cdot k = i \cdot k \leqslant |i|\,|k| \leqslant |i|\,|j| \leqslant d_P d_Q$ for any $k \in \mathbb{N}^n$ with $Q_k \neq 0$. The case when $w = j$ is handled similarly. □

Now consider $\hat{P} := \tau_w(P)$, $\hat{Q} := \tau_w(Q)$, and $\hat{G} = \gcd(\hat{P}, \hat{Q})$ in $\mathbb{K}[x_1, x_1^{-1}, \ldots, x_n, x_n^{-1}, t, t^{-1}]$. We normalize $\hat{G}$ in such a way that $\mathrm{val}_t \hat{G} = 0$ and such that $\hat{G}$ is monic as a polynomial in $t$; this is always possible since $\hat{G}$ is $w$-regular. Let $\alpha = (\alpha_1, \ldots, \alpha_n) \in \mathbb{K}^n$ be an admissible ratio or an FFT ratio. For any $i \in \mathbb{N}$, we evaluate at $x_k := \alpha_k^i$ and $t := t$, which leads us to define the univariate polynomials

$$\begin{aligned} \hat{P}^{[i]} &:= \hat{P}(\alpha_1^i, \ldots, \alpha_n^i, t) \\ \hat{Q}^{[i]} &:= \hat{Q}(\alpha_1^i, \ldots, \alpha_n^i, t) \\ \hat{G}^{[i]} &:= \hat{G}(\alpha_1^i, \ldots, \alpha_n^i, t). \end{aligned}$$

With high probability, $\mathrm{ec}_t \hat{G}^{[i]} = \mathrm{ec}_t \hat{G}$, and $\hat{G}^{[i]}$ is the monic gcd of $\hat{P}^{[i]} t^{-k}$ and $\hat{Q}^{[i]} t^{-\ell}$, where $k := \mathrm{val}_t \hat{P}^{[i]}$ and $\ell := \mathrm{val}_t \hat{Q}^{[i]}$. For sufficiently large $m$ (with $m = O(s)$), we may thus recover $\hat{G}$ from $\hat{G}^{[1]}, \ldots, \hat{G}^{[m]}$ using sparse interpolation. Finally, for $t := 1$, we obtain $G = x_1^{\nu_1} \cdots x_n^{\nu_n} \hat{G}(x_1, \ldots, x_n, 1)$, where $\nu_k = \min(\mathrm{val}_{x_k} P, \mathrm{val}_{x_k} Q) - \mathrm{val}_{x_k} \hat{G}$ for $k = 1, \ldots, n$. This leads to the following algorithm:

**Algorithm 3.2**
**Input:** $P, Q \in \mathbb{K}[x_1, \ldots, x_n]^{\neq}$
**Output:** $G \in \mathbb{K}[x_1, \ldots, x_n]$, such that $G = \gcd(P, Q)$

Find a regularizing weight $w$ for $P$ or $Q$
For $m = 1, 2, 4, 8, \ldots$ do
  Compute $\hat{P}^{[i]}$, $\hat{Q}^{[i]}$ for $i = \lfloor m/2 \rfloor + 1, \ldots, m$
  Compute $\hat{G}^{[i]} = \gcd(\hat{P}^{[i]}, \hat{Q}^{[i]})$ with $\hat{G}^{[i]}$ monic and $\operatorname{val}_t \hat{G}^{[i]} = 0$ for $i = \lfloor m/2 \rfloor + 1, \ldots, m$
  If $\hat{G}^{[1]}, \ldots, \hat{G}^{[m]}$ yield $\hat{G}$ through sparse interpolation, then
    Let $\nu_k := \min(\operatorname{val}_{x_k} P, \operatorname{val}_{x_k} Q) - \operatorname{val}_{x_k} \hat{G}$ for $k = 1, \ldots, n$
    Return $x^{\nu} \hat{G}(x_1, \ldots, x_n, 1)$

**Remark 3.7.** If $\hat{G}$ is normalized in $\mathbb{K}[x_1, x_1^{-1}, \ldots, x_n, x_n^{-1}, t, t^{-1}]$ to be monic as a polynomial in $t$, then we may need to interpolate multivariate polynomials with negative exponents in order to recover $\hat{G}$ from $\hat{G}^{[1]}, \ldots, \hat{G}^{[m]}$. In practice, many interpolation algorithms based on geometric sequences, like Prony's method, can be adapted to do this.

As in the previous subsection, assume that $\alpha_1, \ldots, \alpha_n$ are independently chosen at random from a subset of $\mathbb{K}^{\neq}$ of size at least $N$. We let $d := \max(d_P, d_Q)$, $s := s_G$, and $\bar{s} := s_P + s_Q + s_G$.

LEMMA 3.8. *Assume that $P$ or $Q$ is $w$-regular. Let $\hat{G} = \gcd(\hat{P}, \hat{Q})$ with $\operatorname{val}_t \hat{G} = 0$ be monic as a polynomial in $t$. Take $\hat{P}^{[i]} := \hat{P}(\alpha_1^i, \ldots, \alpha_n^i, t)$, $\hat{Q}^{[i]} := \hat{Q}(\alpha_1^i, \ldots, \alpha_n^i, t)$, and $\hat{G}^{[i]} := \hat{G}(\alpha_1^i, \ldots, \alpha_n^i, t)$. The probability that $\hat{G}^{[i]} = \gcd(\hat{P}^{[i]}, \hat{Q}^{[i]})$ for all $i = 1, \ldots, s$ is at least $1 - 2d^2 \binom{s}{2} / N$.*

**Proof.** We have $\hat{G}^{[i]} = \gcd(\hat{P}^{[i]}, \hat{Q}^{[i]})$ if and only if $R := \operatorname{Res}_t(\hat{P} / \hat{G}, \hat{Q} / \hat{G})$ does not vanish at $\alpha^i$. Now the degree of $R$ is at most $2d^2$, so the probability that $R(\alpha^i) \neq 0$ for a randomly chosen $\alpha \in \mathbb{K}^{\neq}$ and all $i \in \{1, \ldots, s\}$ is at least $1 - 2d^2 \binom{s}{2} / N$, by Corollary 2.2. □

COROLLARY 3.9. *The probability that one can recover $\hat{G}$ from $\hat{G}^{[1]}, \ldots, \hat{G}^{[m]}$ with $m = O(s)$ using sparse interpolation is at least $1 - O(d^2 s^2 / N)$.* □

THEOREM 3.10. *Let $s := s_G$, $\bar{s} := s_P + s_Q + s_G$, $d := \max(d_P, d_Q)$, and $e := \max(\operatorname{ec}_w P, \operatorname{ec}_w Q) \leqslant d^2$. Algorithm 3.2 is correct with probability at least $1 - O(d^2 s^2 / N)$ and it runs in time*

$$O((\bar{s} / s + e)\, \mathsf{S}(s) + s\, \mathsf{M}(e) \log e). \tag{3.1}$$

**Proof.** The correctness with the announced probability follows from Corollaries 2.11 and 3.9, while also using Remark 2.5. The computation of $\hat{P}^{[i]}$ and $\hat{Q}^{[i]}$ through sparse evaluation at geometric progressions requires $O((\bar{s} / s + e)\, \mathsf{S}(s))$ operations (see Section 2.4). The univariate gcd computations take $O(s\, \mathsf{M}(e) \log e)$ further operations. The final interpolation of $\hat{G}$ from $\hat{G}^{[1]}, \ldots, \hat{G}^{[m]}$ can be done in time $O(\mathsf{S}(s))$. □

**Example 3.11.** Let $P, Q \in \mathbb{K}[x_1, \ldots, x_n]^{\neq}$. Consider the particular case when $P$ is monic as a polynomial in $\mathbb{K}[x_2, \ldots, x_n][x_1]$. Then, $w = (1, 0, \ldots, 0)$ is a regularizing weight for $P$, and therefore also for $G := \gcd(P, Q)$. This situation can be readily detected and, in this case, we have $e \leqslant \max(\deg_{x_1} P, \deg_{x_1} Q)$ in the complexity bound (3.1).

**Remark 3.12.** We may need fewer evaluation points to interpolate the gcd $\hat{G}$ in Algorithm 3.2 in case the terms of $\hat{G}$ are distributed over the powers of $t$. For instance, if the terms are distributed evenly, then we have $s := s_G / e$ in the complexity bound (3.1).

# 4. BIVARIATE FACTORIZATION

Lemma 3.2 allows us to project the general problem of multivariate gcd computations down to the univariate case. For the polynomial factorization problem, no such reduction exists: given a random univariate polynomial of degree $d \geqslant 2$ over a finite field, there is a non-zero probability that this polynomial is not irreducible. For this reason, it is customary to project down to bivariate instead of the univariate polynomials (when applicable, an alternative is to project down to univariate polynomials with integer coefficients; see [79], for instance).

This explains why it is interesting to study the factorization of bivariate polynomials in more detail. Throughout this section, $F$ is a bivariate polynomial in $\mathbb{K}[x,y]^{\neq}$ of degree $d_x$ in $x$ and of degree $d_y$ in $y$. As in Sections 2.2 and 3, random numbers will be drawn from a subset of $\mathbb{K}^{\neq}$ of size at least $N$. We will recall some classical results concerning the complexity of bivariate factorization, important techniques, and special cases: content-free factorization, root extraction, Hensel lifting, and square-free factorization.

## 4.1. Content-free factorization

Recall that the *content* of $F = F_0 + \cdots + F_{d_x} x^{d_x} \in \mathbb{K}[x,y]^{\neq}$ in $x$ is defined by

$$\operatorname{cont}_x F \;:=\; \gcd(F_0, \ldots, F_{d_x}) \;\in\; \mathbb{K}[y].$$

We say that $F$ is *content-free* (or *primitive*) in $x$ if $\operatorname{cont}_x F = 1$. Given two random shifts $\sigma, \tau \in \mathbb{K}^{\neq}$, we have $\operatorname{cont}_x F = \gcd(F(\sigma,y), F(\tau,y))$ with high probability. More precisely:

PROPOSITION 4.1. *The content* $\operatorname{cont}_x F$ *can be computed in time* $O(d_x d_y + \mathsf{M}(d_y) \log d_y)$ *with a probability of success of at least* $1 - 2 d_x^2 / N$.

**Proof.** Without loss of generality, we may assume that $|\mathbb{K}| > N > 2 d_x^2$. Let us first consider the case when $\operatorname{cont}_x F = 1$. Then we claim that $\gcd(F(\sigma,y) : \sigma \in \mathbb{K}^{\neq}) = 1$. Indeed, for $d_x + 1$ pairwise distinct $\sigma_0, \ldots, \sigma_{d_x} \in \mathbb{K}^{\neq}$, the Vandermonde matrix $(\sigma_i^j)_{0 \leqslant i,j \leqslant d_x}$ is invertible, so $\gcd(F(\sigma_0,y), \ldots, F(\sigma_{d_x},y)) = \gcd(F_0, \ldots, F_{d_x}) = 1$. It follows that $\operatorname{Res}_y(F(u,y), F(v,y)) \neq 0$, regarded as an element of $\mathbb{K}[u,v]$, is non-zero, and its total degree is bounded by $2 d_x^2$. By Lemma 2.1, it follows that $\operatorname{Res}_y(F(\sigma,y), F(\tau,y)) \neq 0$ with probability at least $1 - 2 d_x^2 / N$. In that case, $\gcd(F(\sigma,y), F(\tau,y)) = 1$.

We have proved our probabilistic correctness claim in the particular case when $\operatorname{cont}_x F = 1$. In general, we factor $F = \tilde{F} \operatorname{cont}_x F$. With probability at least $1 - 2 d_x^2 / N$, we have $\gcd(F(\sigma,y), F(\tau,y)) = \gcd(\tilde{F}(\sigma,y), \tilde{F}(\tau,y)) \operatorname{cont}_x F = 1$.

As to the complexity bound, the evaluations $F(\sigma,y)$ and $F(\tau,y)$ require $O(d_x d_y)$ operations and the univariate gcd computation can be done in time $\mathsf{M}(d_y) \log d_y$. □

## 4.2. Root extraction

Let $F \in \mathbb{K}[x,y]^{\neq}$ and $\ell \geqslant 2$. Assume that $F = c R^{\ell}$ for some $c \in \mathbb{K}^{\neq}$ and $R \in \mathbb{K}[x,y]$. Assume that $|\mathbb{K}| > N > \binom{d_y}{2} + 2$. Then $R$ can be computed efficiently as follows.

After dividing out a suitable power of $x^\ell$, we may assume without loss of generality that $\mathrm{val}_x F = 0$. For a random shift $\sigma$, we next replace $F(x,y)$ with $F(x,y+\sigma)$. With high probability, this ensures that $F(0,0) \neq 0$. Modulo division of $F(x,y)$ by $F(0,0)$, we may then assume without loss of generality that $F(0,0)=1$, and we will show how to compute the unique $R \in \mathbb{K}[x,y]$ with $R(0,0)=1$ and $R^\ell = F$. Since $F(0,y) \in \mathbb{K}[y]$ is a univariate polynomial, we may efficiently compute $R(0,y)$ using univariate root extraction.

Let $\alpha \in \mathbb{K}^{\neq}$ be an admissible ratio or an FFT ratio. For any $i \in \mathbb{N}$, we define the univariate polynomial $F^{[i]} := F(x, \alpha^i)$. With high probability, we have $F^{[i]}(0) \neq 0$. Let $R^{[i]}$ be the unique univariate polynomial such that $(R^{[i]})^\ell = F^{[i]}$ and $R^{[i]}(0) = R(0, \alpha^i)$. Such $R^{[i]}$ can be found efficiently using univariate root extraction. For $m = d_y$, we may recover $R$ from $R^{[0]}, \ldots, R^{[m]}$ using interpolation.

PROPOSITION 4.2. *With the above notations and assumptions, we may compute c and R in time* $O(\mathsf{M}(d_x d_y))$, *with a probability of success of at least* $1 - \left(\binom{d_y}{2} + 2 d_y\right)/N$.

**Proof.** The random shift $F(x,y) \mapsto F(x,y+\sigma)$ and the corresponding backshift $R(x,y) \mapsto R(x,y-\sigma)$ can be computed in time $O(d_x \mathsf{M}(d_y))$ using the so-called convolution method [2, Theorem 5]. The computation of $R(0,y)$ for $i=0,\ldots,d_y$ can be done in time $O(\mathsf{M}(d_y))$ using fast root extraction. We may compute $R(0,\alpha^i)$ and $F^{[i]}$ for $i=0,\ldots,d_y$ in time $O(d_x \mathsf{M}(d_y))$ using fast multipoint evaluation at geometric sequences [2, 13]. The same holds for the interpolation of $R$ from the $R^{[i]}$. Finally, the $\ell$-th roots $R^{[i]}$ of the univariate polynomials $F^{[i]}$ can be computed in time $O(d_y \mathsf{M}(d_x))$. The algorithm is correct as long as $F(0,0) \neq 0$ and $F(0,\alpha^i) \neq 0$ for $i=0,\ldots,d_y$. The probability that $F(0,0) \neq 0$ and $F(0,1) \neq 0$ for a random choice of $\sigma$ is at least $1 - 2 d_y / N$, by Lemma 2.1. In that case, $\Phi := F(0,y) \neq 0$ and the probability that $\Phi(\alpha^i) \neq 0$ for $i=1,\ldots,d_y$ is at least $1 - \binom{d_y}{2}/N$, by Corollary 2.2. □

## 4.3. Hensel lifting

Let $F \in \mathbb{K}[x,y]$ be content-free in $x$ and assume that $F$ has a non-trivial factorization $F = PQ$, and $\mathrm{cont}_x F = 1$. Assume that $\deg F(x,0) = d_x$ and that $P(x,0)$ and $Q(x,0)$ are known and coprime (in particular $P$ and $Q$ are coprime). Using a random shift $F(x,y) \mapsto F(x,y+\sigma)$, these assumptions can be enforced with high probability, provided that $P$ and $Q$ are coprime. Without loss of generality we may also assume that we normalized the factorization of $F(x,0)$ such that $P(x,0)$ is monic. Under these assumptions, we may compute $P$ and $Q$ as follows:

- We first use Hensel lifting to obtain a factorization $F = \hat{P}\hat{Q}$ with $\hat{P}(x,y), \hat{Q}(x,y) \in \mathbb{K}[[y]][x]$ and $\hat{P}(x,0) = P(x,0)$, $\hat{Q}(x,0) = Q(x,0)$, and such that $\hat{P}$ is monic in $x$. We compute $\hat{P}$ and $\hat{Q}$ modulo $y^\nu$ for $\nu = 2 d_y + 1$.
- For a random shift $\sigma \in \mathbb{K}^{\neq}$, we apply rational function reconstruction [31, Section 5.7] to $\hat{P}(\sigma, y)$ to obtain $A, B \in \mathbb{K}[y]$ with $\hat{P}(\sigma, y) = A/B + O(y^\nu)$ and $\gcd(A,B) = 1$ and $B$ of the smallest possible degree with these properties. With high probability, we then have $P = B\hat{P}$. We may compute $Q$ in a similar way.

PROPOSITION 4.3. *Given F and* $F(x,0) = P(x,0)\, Q(x,0)$ *satisfying the above assumptions, let* $d_x = \deg_x F$, $d_y = \deg_y F$, *and* $\delta := \max(d_x, d_y)$. *We may lift the factorization of* $F(x,0)$ *into a factorization* $F = PQ$ *in time*

$$O(\mathsf{M}(d_x d_y) + \mathsf{M}(\delta) \log \delta),$$

*with a probability of success of at least* $1 - 2 d_x d_y / N$.

**Proof.** We first observe that there is a unique factorization $F=PQ$ that lifts the factorization of $F(x,0)$, thanks to our hypothesis that $\mathrm{cont}_x F=1$. Since this hypothesis also implies that $\mathrm{cont}_x P=\mathrm{cont}_x Q=1$, the denominator $B$ of $\hat{P}$ as an element of $\mathbb{K}(y)[x]$ coincides with the leading coefficient of $P$ as a polynomial in $x$. Consequently, the denominator of $\hat{P}(\sigma,y)$ equals $B$ if and only if $\mathrm{Res}_y(B(y),P(\sigma,y))\neq 0$. Since the degree of $\mathrm{Res}_y(B(y),P(u,y))\in\mathbb{K}[u]$ is bounded by $d_x d_y$, this happens with probability at least $1-d_x d_y/N$. Since the degrees of the numerator $A$ and denominator $B$ of $\hat{P}(\sigma,y)\in\mathbb{K}(y)$ do not exceed $d_y$, it suffices to compute $\hat{P}$ modulo $O(y^{2d_y+1})$ in order to recover $A$ and $B$. This completes the probabilistic correctness proof.

As to the complexity bound, the Hensel lifting requires $O(\mathsf{M}(d_x d_y)+\mathsf{M}(d_x)\log d_x)$ operations in $\mathbb{K}$, when using a fast Newton iteration [31, Theorem 15.18, two factors]. The computation of $\hat{P}(\sigma,y)$ requires $O(d_x d_y)$ further operations and the rational function reconstruction can be done in time $O(\mathsf{M}(d_y)\log d_y)$ using the technique of half gcds [31, Chapter 11]. □

The proposition generalizes in a straightforward way to the case when $F$ has $\ell$ pairwise coprime factors $P_1,\ldots,P_\ell$. In that case, one may use fast multifactor Hensel lifting [31, Theorem 15.18], to obtain the following:

PROPOSITION 4.4. *Let $F\in\mathbb{K}[x,y]$ be such that $\mathrm{cont}_x F=1$ and $\deg F(x,0)=d_x$. Assume that $F$ can be factored as $F=P_1\cdots P_\ell$, where $P_1(x,0),\ldots,P_\ell(x,0)$ are pairwise coprime and known. Assume also that $P_1(x,0),\ldots,P_{\ell-1}(x,0)$ are monic. Then we may lift the factorization $F(x,0)=P_1(x,0)\cdots P_\ell(x,0)$ into a factorization $F=P_1\cdots P_\ell$ in time*

$$O(\mathsf{M}(d_x d_y)\log \ell+\ell\,\mathsf{M}(\delta)\log\delta),$$

*with a probability of success of at least $1-\ell d_x d_y/N$.* □

**Example 4.5.** Consider

$$F \;:=\; x^3y^2-x^3+x^2y^3+x^2+xy^2+3xy-2x+2y^2-2y$$

with

$$F(x,0) \;=\; -x^3+x^2-2x \;=\; (-x^2+x-2)\,x.$$

This factorization lifts to the following factorization of $F$ in $\mathbb{Q}[[y]][x]$:

$$F \;=\; \hat{P}\hat{Q}, \qquad \hat{P} \;=\; x^2+\frac{x}{y-1}+\frac{2}{y+1}, \qquad \hat{Q} \;=\; (y^2-1)\,x+y^3-y.$$

Taking $\sigma=1$, we obtain the following rational reconstruction of $\hat{P}(\sigma,y)$ up to the order $O(y^7)$:

$$\hat{P}(1,y) \;=\; 1+\frac{1}{y-1}+\frac{2}{y+1} \;=\; \frac{y^2+3y-2}{y^2-1}.$$

Consequently, $P=(y^2-1)\,\hat{P}$ is the sought factor of $F$ in $\mathbb{Q}[x,y]$. In a similar way, we find that $Q=(y^2-1)^{-1}\,\hat{Q}$.

## 4.4. Square-free factorization

Assume that $F\in\mathbb{K}[x,y]$ is content-free in $y$ and of total degree $d$. Assume also that $\mathrm{char}\,\mathbb{K}>d$. Recall that the *square-free factorization* of $F$ is of the form

$$F \;=\; P_1^1 P_2^2\cdots P_d^d,$$

where each $P_i$ is the product of all irreducible factors of $F$ that occur with multiplicity $i$. Note that some of the $P_i$ are allowed to be units in $\mathbb{K}$ and that the $P_i$ are unique up to multiplication by such units. The polynomials $P_1, \ldots, P_d$ are pairwise coprime. Since $\operatorname{char} \mathbb{K} > d$, they must also be separable in both $x$ and $y$ (i.e. gcd $(P_i, \partial P_i / \partial x) = \gcd (P_i, \partial P_i / \partial y) = 1)$. The square-free factorization of $F = \mathbb{K}[x, y]$ can be computed efficiently as follows:

- For a random shift $\sigma$, replace $F$ by $F(x, y + \sigma)$.
- Compute the square-free factorization of $F(x, 0)$.
- Hensel lift this into the square-free factorization of $F$ using Proposition 4.4.
- Apply the shift in the opposite direction.

PROPOSITION 4.6. *Assume that $F \in \mathbb{K}[x, y]$ is content-free in $y$ and that* $\operatorname{char} \mathbb{K} > d$. *We can compute the square-free factorization of $F$ in time*

$$O(\mathsf{M}(d_x d_y) \log \ell + \mathsf{M}(d_y)\, \ell \log d_y + \mathsf{M}(d_x) \log d_x),$$

*with a probability of success of at least* $1 - 3\, \ell d_x d_y / N$, *where* $\ell := |\{1 \leqslant i \leqslant d : P_i \notin \mathbb{K}\}|$.

**Proof.** Given $i \in \{1, \ldots, d\}$, consider $P_i$ and $\bar{P}_i := (\partial P_i / \partial x) F / P_i$. The polynomials $P_i(x, \sigma)$ and $\bar{P}_i(x, \sigma)$ are coprime if and only if $\operatorname{Res}_x(P_i(x, u), \bar{P}_i(x, u)) \in \mathbb{K}[u]$ does not vanish at $u := \sigma$. Since this resultant has degree at most $2 d_x d_y$, this happens with probability $1 - 2 d_x d_y / N$. Therefore, the probability that all $P_i(x, \sigma)$ are pairwise coprime and all $P_i(x, \sigma)$ are separable is at least $1 - 2\, \ell d_x d_y / N$. In that case, $F(x, \sigma) = P_1(x, \sigma)\, P_2(x, \sigma)^2 \cdots P_d(x, \sigma)^d$ is the square-free decomposition of $F(x, \sigma)$. Modulo normalization, we are thus in the position to apply Proposition 4.4. This proves the probabilistic correctness of the algorithm.

The computation of the shift $F(x, y) \mapsto F(x, y + \sigma)$ can be done in time $O(d_x \mathsf{M}(d_y))$ using the algorithm from [2, Theorem 5] and the same holds for the shifts in the opposite direction in the last step. The square-free factorization of the univariate polynomial $F(x, 0)$ can be done in time $O(\mathsf{M}(d_x) \log d_x)$: see [106] and [31, Theorem 14.23]. We conclude with Proposition 4.4. □

## 4.5. General bivariate factorization

General bivariate factorization of $F \in \mathbb{K}[x, y]$ can often be reduced to Hensel lifting as in Section 4.3 using a random shift $y \mapsto y + \sigma$ and diversification $x \mapsto \zeta_1 x$, $y \mapsto \zeta_2 y$. Let $d_x = \deg_x F$, $d_y = \deg_y F$. The best currently known complexity bound is the following:

THEOREM 4.7. [74, Proposition 8] *Let $F \in \mathbb{K}[x, y]$ be square-free and content-free in both $x$ and $y$. Assume that* $\operatorname{char} \mathbb{K} = 0$ *or* $\operatorname{char} \mathbb{K} > d_y (2 d_x - 1)$. *Then we can compute the irreducible factorization of $F$ in time*

$$\tilde{O}(d_x^2 d_y + d_x^{\omega}) + \mathsf{F}(d_x)$$

*and with a probability of success of at least* $1 - d_x / N$. □

The actual proposition in [74] also contains a similar result for finite fields of small characteristic. For $\mathbb{K}$ as in Theorem 4.7, square-freeness and content-freeness can be achieved with high probability and negligible cost using the algorithms from Sections 4.1 and 4.4.

# 5. EFFICIENT REDUCTIONS

In the bivariate setting of the previous section, we have presented several efficient algorithms for the computation of partial factorizations. In this section, we will generalize three of them to the multivariate setting: removal of content, root extraction, and squarefree factorizations. The common feature of these generalizations is that they recover the answer directly from the corresponding univariate specializations of the problem, in a similar fashion as the gcd algorithm from Section 3.2.

## 5.1. Content-free factorization

Consider a polynomial $F \in \mathbb{K}[x_1, \ldots, x_n] \setminus \mathbb{K}$ and a variable $x_i$. If, for every non-trivial factorization $F = P\,Q$ with $P, Q \in \mathbb{K}[x_1, \ldots, x_n] \setminus \mathbb{K}$, both $P$ and $Q$ depend on $x_i$, then we say that $F$ is *content-free* (or *primitive*) in $x_i$. Note that this is always the case if $\deg_{x_i} F \leqslant 1$. If $F$ is content-free in $x_i$ for all $i = 1, \ldots, n$, then we say that $F$ is *content-free*.

For a given variable $x_i$, say $x_1$, we can efficiently test whether $F$ is content-free with respect to $x_i$: for random $\alpha_2, \ldots, \alpha_n \in \mathbb{K}^{\neq}$, we form the bivariate polynomial $B := F(x_1, \alpha_2 t, \ldots, \alpha_n t)$ and compute $\operatorname{cont}_{x_1} B \in \mathbb{K}[t]$ using the method from Section 4.1. With high probability, $F$ is content-free with respect to $x_1$ if and only if $\operatorname{cont}_{x_1} B = 1$. Performing this test for each of the variables $x_1, \ldots, x_n$ yields:

PROPOSITION 5.1. *We may check whether $F$ is content-free (and, if not, determine all variables $x_i$ with respect to which $F$ is not content-free) in time $O(n\,s_F + n\,\mathsf{M}(d_F) \log d_F)$ and with a probability of success of at least $1 - 2\,n\,d_F^2 / N$.*

**Proof.** The proof is similar to the one of Proposition 4.1. This time, for the probability bound, we consider the resultant $\operatorname{Res}_t(F(u, c_2 t, \ldots, c_n t), F(v, c_2 t, \ldots, c_n t))$ as a polynomial in $\mathbb{K}[u, v, c_2, \ldots, c_n]$, of total degree at most $2\,d_F^2$. If $\operatorname{cont}_{x_1} F = 1$, then this resultant does not vanish with probability at least $1 - 2\,d_F^2 / N$ for random $u := \sigma$, $v := \tau$, $c_2 := \alpha_2, \ldots, c_n := \alpha_n$.

As to the complexity bound, for $i = 1, \ldots, n$, let $\sigma_i$ and $\tau_i$ be random and consider $B_i := F(\alpha_1 t, \ldots, \alpha_{i-1} t, \sigma_i, \alpha_{i+1} t, \ldots, \alpha_n t)$ and $C_i := F(\alpha_1 t, \ldots, \alpha_{i-1} t, \tau_i, \alpha_{i+1} t, \ldots, \alpha_n t)$. We compute the $B_i$ simultaneously for $i = 1, \ldots, n$ in time $O(n\,(s_F + d_F))$ and similarly for the $C_i$. Finally, the computation of $\gcd(B_i, C_i)$ for $i = 1, \ldots, n$ takes $O(n\,\mathsf{M}(d_F) \log d_F)$ operations. □

Assume now that $F$ is not content-free, say with respect to $x_1$. With high probability, the content of $F$ with respect to $x_1$ then equals the gcd of $F(\sigma, x_2, \ldots, x_n)$ and $F(\tau, x_2, \ldots, x_n)$, for two random shifts $\sigma, \tau \in \mathbb{K}$. This leads to a non-trivial factorization of $F$ for the cost of one gcd computation and one exact division.

## 5.2. Root extraction

Given $F \in \mathbb{K}[x_1, \ldots, x_n]^{\neq}$ and $\ell \geqslant 2$, multivariate $\ell$-th root extraction is the problem of determining $c \in \mathbb{K}^{\neq}$ and $R \in \mathbb{K}[x_1, \ldots, x_n]$, such that $F = c\,R^{\ell}$, whenever such $c$ and $R$ exist. We devise an algorithm in the same vein as the algorithm for gcd computations from Section 3.2.

We first compute a regularizing weight $w$ for $F$ such that $\operatorname{ec}_w F$ is small. Recall that the regularity of $w$ ensures that $\operatorname{lp}_w F = \alpha\,x^{\nu}$ for some $\alpha \in \mathbb{K}^{\neq}$ and $\nu \in \mathbb{Z}^n$. We take $c := \alpha$ and note that we then must have $\operatorname{lp}_w F = c\,(\operatorname{lp}_w R)^{\ell}$.

Now let $\hat{F} = (\mathrm{lp}_w F)^{-1} t^{-\mu} \tau_w(F) \in \mathbb{K}[x_1, x_1^{-1}, \ldots, x_n, x_n^{-1}, t, t^{-1}]$ with $\mu := \mathrm{val}_t \tau_w(F)$, so that $\mathrm{val}_t \hat{F} = 0$ and $\hat{F}$ is monic as a polynomial in $t$. Let $\alpha = (\alpha_1, \ldots, \alpha_n) \in \mathbb{K}^n$ be an admissible ratio or an FFT ratio. For any $i \in \mathbb{N}$, we define the univariate polynomial $\hat{F}^{[i]} := \hat{F}(\alpha_1^i, \ldots, \alpha_n^i, t)$. Let $\hat{R}^{[i]}$ be the unique monic polynomial with $(\hat{R}^{[i]})^\ell = \hat{F}^{[i]}$. For sufficiently large $m$ (with $m = O(s_R)$), we may recover $\hat{R}$ from $\hat{R}^{[1]}, \ldots, \hat{R}^{[m]}$ using sparse interpolation. Finally, we have $R = x^\nu \hat{R}(x_1, \ldots, x_n, 1)$, where $\nu_i \in \mathbb{Z}$ is such that $\nu_i^\ell = \mathrm{val}_{x_i} F$ for $i = 1, \ldots, n$.

PROPOSITION 5.2. *Assume that $F$ is $w$-regular with $e := \mathrm{ec}_w F \leqslant d_F^2$. Then we may compute $c \in \mathbb{K}^{\neq}$ and $R \in \mathbb{K}[x_1, \ldots, x_n]$ with $F = c R^\ell$, whenever such $c$ and $R$ exist, in time*

$$O((s_F / s_R + e)\, \mathsf{S}(s_R) + s_R \mathsf{M}(e)).$$

**Proof.** The evaluations $\hat{F}^{[i]} := \hat{F}(\alpha_1^i, \ldots, \alpha_n^i, t)$ take $O((s_F / s_R + e)\, \mathsf{S}(s_R))$ operations, whereas the sparse interpolation of $\hat{R}$ from the $\hat{R}^{[i]}$ can be done in time $O(\mathsf{S}(s_R))$. The cost of the univariate $\ell$-th root extractions $\hat{R}^{[i]} := \sqrt[\ell]{\hat{F}^{[i]}}$ is bounded by $O(s_R \mathsf{M}(e))$. □

## 5.3. Square-free factorization

Consider $F \in \mathbb{K}[x_1, \ldots, x_n]$ of total degree $d$. Assume that $F$ is content-free and that $\mathrm{char}\, \mathbb{K} = 0$ or $\mathrm{char}\, \mathbb{K} > d^2$. The factorization of

$$F = c P_1 P_2^2 \cdots P_d^d$$

into square-free parts can be done using a similar technique as for gcd computations in Section 3.2. We start with the computation of a regularizing weight $w$ for $F$. Setting $e := \mathrm{ec}_w F$, we recall that $e \leqslant d^2$, whence $\mathrm{char}\, \mathbb{K} > e$. Let

$$\hat{F} = \tau_w(F) \in \mathbb{L} := \mathbb{K}[x_1, x_1^{-1}, \ldots, x_n, x_n^{-1}, t, t^{-1}]$$

and consider the normalized square-free factorization

$$\hat{F} = c x^\nu t^\mu \hat{P}_1 \hat{P}_2^2 \cdots \hat{P}_d^d,$$

where $c \in \mathbb{K}^{\neq}$, $\nu \in \mathbb{Z}^n$, $\mu \in \mathbb{Z}$, and where $\hat{P}_1, \ldots, \hat{P}_d \in \mathbb{L}$ are monic and of valuation zero in $t$. Let $\alpha = (\alpha_1, \ldots, \alpha_n) \in \mathbb{K}^n$ be an admissible ratio or an FFT ratio. For any $i \in \mathbb{N}$, we define the univariate polynomials

$$\begin{aligned} \hat{F}^{[i]} &:= \hat{F}(\alpha_1^i, \ldots, \alpha_n^i, t) \\ \hat{P}_k^{[i]} &:= \hat{P}_k(\alpha_1^i, \ldots, \alpha_n^i, t), \qquad k = 1, \ldots, d. \end{aligned}$$

The normalized square-free factorization of $\hat{F}^{[i]} t^{-\mu}$ is of the form

$$\hat{F}^{[i]} t^{-\mu} = c^{[i]} \hat{P}_1^{[i]} (\hat{P}_2^{[i]})^2 \cdots (\hat{P}_d^{[i]})^d,$$

where $c^{[i]} \in \mathbb{K}^{\neq}$, and $\hat{P}_1^{[i]}, \ldots, \hat{P}_d^{[i]}$ are monic polynomials in $\mathbb{K}[t]$. We recover $\hat{P}_1, \ldots, \hat{P}_d$ using sparse interpolation and then $P_1, \ldots, P_d$ by setting $t := 1$ and multiplying by appropriate monomials in $x_1^{\mathbb{Z}} \cdots x_n^{\mathbb{Z}}$. More precisely, $P_i = x^\lambda \hat{P}_i(x_1, \ldots, x_n, 1)$, where $\lambda_j = 1$ if $\mathrm{val}_{x_j} F = i$ and $\lambda_j = 0$ otherwise.

PROPOSITION 5.3. *Assume that $F$ is $w$-regular with $e := \mathrm{ec}_w F \leqslant d_F^2$. Let $s := \max(s_{P_1}, \ldots, s_{P_d})$ and $\ell := |\{1 \leqslant i \leqslant d : P_i \notin \mathbb{K}\}|$. Then we may compute a square-free factorization of in time*

$$O(\ell\, (s_F / s + e)\, \mathsf{S}(s) + s \mathsf{M}(e) \log e),$$

*with a probability of success of at least* $1-3\,\ell d^2 s^2/N$.

**Proof.** The probabilistic correctness is proved in a similar way as in the bivariate case (see Proposition 4.6), while also using Corollary 2.2. The sparse evaluation and interpolation at a geometric sequence require $O((s_F/s_{P_1}+e)\,\mathsf{S}(s_{P_1})+\cdots+(s_F/s_{P_d}+e)\,\mathsf{S}(s_{P_d}))=O(\ell\,(s_F/s+e)\,\mathsf{S}(s))$ operations. The univariate square-free factorizations can be done in time $O(s\,\mathsf{M}(e)\log e)$, using [106] and [31, Theorem 14.23]. □

## 5.4. A pathological example

It is well known that a divisor of a sparse polynomial can have far more terms than the polynomial itself, because of the identity

$$x^k-1 \;=\; (x-1)\,(x^{k-1}+\cdots+1), \tag{5.1}$$

and the possibility to replace $x$ by any other sparse polynomial. For this reason, many problems on sparse polynomials turn out to be NP-hard [89, 90]. In a way that has been made precise in [25], this fundamental example is actually the main cause for such hardness results.

The example (5.1) is less dramatic if we consider sparse polynomials for which the total degree is much smaller than the total number of terms, which is often the case in practice. But even then, it still has some unpleasant consequences. Recall from [25] that

$$\gcd\,(x^{pq}-1, x^{p+q}-x^p-x^q+1) \;=\; x^{p+q-1}+\cdots+x^p-x^{q-1}-\cdots-1$$

for coprime $p$, $q$ with $p>q\geqslant 4$. This gcd contains exactly $2q$ terms. Such gcds can also occur during content-free factorizations. For instance, the content of

$$F_{p,q}(x,y) \;=\; x^{pq}-1+(x^{p+q}-x^p-x^q+1)\,y$$

in $y$ is $x^{p+q-1}+\cdots+x^p+x^{q-1}+\cdots+1$. Now consider

$$F \;\coloneqq\; F_{p,q}(x_1,y)\cdots P_{p,q}(x_n,y).$$

Then $\deg F=n p q$ and $s_F\leqslant 6^n$. The size of each individual factor in any irreducible factorization of $F$ is bounded by $p q$, which is sharp with respect to $\deg F$. However, the content $C$ of $F$ in $y$ has size $s_C=(2q)^n$. This means that content-free factorization as a preparation step (before launching a "more expensive" factorization method) is a bad idea on this particular example.

# 6. FACTORING USING ITERATED HENSEL LIFTING

Let $F\in\mathbb{K}[x_1,...,x_k]\setminus\mathbb{K}$ be a content-free polynomial and recall that any factor of $F$ is again content-free. Let $c_1,\ldots,c_n$ be random non-zero elements of $\mathbb{K}$. For any $A\in\mathbb{K}[x_1,\ldots,x_n]$ and $k=0,\ldots,n$, we define

$$A^{[k]}(x_1,\ldots,x_k) \;\coloneqq\; A(x_1,\ldots,x_k,c_{k+1},\ldots,c_n).$$

In this section, we describe several algorithms for the factorization of $F$, following a similar recursive approach as for the computations of gcds in Section 3.1. This time, we start with the computation of a non-trivial factorization of the bivariate polynomial $F^{[2]}$. From this, we will recursively obtain non-trivial factorizations of $F^{[3]},F^{[4]},\ldots,F^{[n]}$. This will in particular yield a non-trivial factorization of $F=F^{[n]}$. We will separately study the cases when $F$ has a factorization into two or more coprime factors.

## 6.1. Faithful projections

In order to reconstruct factorizations of $F$ from factorizations of $F^{[2]}$, it is important that the projection $A \mapsto A^{[2]}$ be sufficiently generic. For $k=0,\ldots,n$, let us denote by $\pi_{c,k}$ the projection $A \mapsto A^{[k]}$. We say that $\pi_{c,k}$ is *faithful* for $A \in \mathbb{K}[x_1,\ldots,x_n]$ if $\operatorname{supp} A^{[l]} = \{(e_1,\ldots,e_l): e \in \operatorname{supp} A\}$ for $l=k,\ldots,n-1$.

As usual, we assume that $c_1,\ldots,c_n$ are independently chosen at random from a subset of $\mathbb{K}^{\neq}$ of size at least $N$.

LEMMA 6.1. *Given $A \in \mathbb{K}[x_1,\ldots,x_n]$, the probability that $\pi_{c,n-1}$ is faithful for $A$ is at least $1-s_A \deg_{x_n} A/N$.*

**Proof.** Given $e \in \operatorname{supp} A$, let $\Phi(x_n) = \sum_{i=0}^{\deg_{x_n} A} A_{e_1,\ldots,e_{n-1},i} x_1^{e_1} \cdots x_{n-1}^{e_{n-1}} x_n^i$. Then $(e_1,\ldots,e_{n-1}) \notin \operatorname{supp} A^{[n-1]}$ if and only if $\Phi(c_n)=0$, which happens with probability at most $\deg_{x_n} A/N$. □

Now consider a factorization $F = P_1 \cdots P_\ell$ such that $P_1,\ldots,P_\ell$ are pairwise coprime. We say that $\pi_{c,k}$ is *faithful* for this factorization if $\pi_{c,k}$ is faithful for $P_1,\ldots,P_\ell$ and $\pi_{c,k}(P_1),\ldots,\pi_{c,k}(P_\ell)$ are pairwise coprime.

LEMMA 6.2. *Given a factorization $F=P_1\cdots P_\ell$ and $k \in \{2,\ldots,n-1\}$, the probability that $\pi_{c,k}$ is faithful for this factorization is at least $1-(n\,\delta_F\,(s_{P_1}+\cdots+s_{P_\ell})+n^2 d_F^2)/N$.*

**Proof.** This directly follows from Lemma 3.2 (while using that $\sum_{i<j} d_{P_i} d_{P_j} < d_F^2$) and the previous lemma (using induction on $n$). □

While faithful mappings preserve coprime factorizations in a predictable way, it may still happen that an irreducible polynomial is projected to a reducible one. In fact this may even happen with probability $1/2$ for a random choice of $c_1,\ldots,c_n$.

**Example 6.3.** Let $n := 3$ and $\mathbb{K} := \mathbb{F}_p$ for an odd prime $p$. Consider

$$
\begin{aligned}
F &:= \Phi(x_1,x_2)^2 - x_3 \\
\Phi &:= x_1 + x_2
\end{aligned}
$$

Then $F$ is irreducible, but $F^{[2]} = (\Phi+\beta)\,(\Phi-\beta)$ whenever $c_3 = \beta^2$ is a square in $\mathbb{K}$. If $c_3$ is a random element of $\mathbb{K}^{\neq}$, then this happens with probability $1/2$. (Note that we may replace $\Phi$ by any other irreducible polynomial that involves both $x_1$ and $x_2$.)

This phenomenon is not so much of a problem if we want to recursively Hensel lift a coprime factorization $F^{[2]} = A\,B$ for which we *know* that there exist $P, Q \in \mathbb{K}[x_1,\ldots,x_n]$ with $A = P^{[2]}$ and $B = Q^{[2]}$ (see Algorithms 6.1 and 6.2 below). However, any algorithm for Hensel lifting will fail if $F$ is irreducible or, more generally, if no such $P$ and $Q$ exist (Algorithm 6.3 below for the irreducible factorization of $F$ may therefore fail on Example 6.3).

## 6.2. Lifting a coprime decomposition into two factors

Consider a non-trivial factorization $F = P\,Q$, where $P, Q \in \mathbb{K}[x_1,\ldots,x_n] \setminus \mathbb{K}$ are coprime. Assume that $\pi_{c,2}$ is faithful for this factorization. Let us show how to recover $P^{[k+1]}$ and $Q^{[k+1]}$ from $P^{[k]}$ and $Q^{[k]}$, for $k=2,\ldots,n-1$.

Let $\alpha = (\alpha_1, \ldots, \alpha_k)$ be an admissible ratio or an FFT ratio. For any $A \in \mathbb{K}[x_1, \ldots, x_n]$ and any $i \in \mathbb{N}$, let

$$A^{\langle k+1,i\rangle}(t,u) \;:=\; A^{[k+1]}(\alpha_1^i t, \ldots, \alpha_k^i t, u)$$

and

$$A^{[k,i]}(t) \;:=\; A^{\langle k+1,i\rangle}(t, c_{k+1}) \;=\; A^{[k]}(\alpha_1^i t, \ldots, \alpha_k^i t).$$

With high probability, we have $\deg_t F^{\langle k+1,i\rangle} = \deg F^{[k]}$, $\deg P^{[k,i]} = \deg P^{[k]}$, $\deg Q^{[k,i]} = \deg Q^{[k]}$, and the polynomials $P^{[k,i]}$ and $Q^{[k,i]}$ are again coprime. It follows that each factorization

$$F^{[k,i]} \;=\; P^{[k,i]} Q^{[k,i]}$$

can be Hensel-lifted (see Section 4.3) into a factorization

$$F^{\langle k+1,i\rangle} \;=\; P^{\langle k+1,i\rangle} Q^{\langle k+1,i\rangle}.$$

We now recover $P^{[k+1]}$ and $Q^{[k+1]}$ from $P^{\langle k+1,1\rangle}, \ldots, P^{\langle k+1,s_{P^{[k]}}\rangle}$ and $Q^{\langle k+1,1\rangle}, \ldots, Q^{\langle k+1,s_{Q^{[k]}}\rangle}$, respectively, using sparse interpolation. In fact, if $s_{P^{[k]}} \leqslant s_{Q^{[k]}}$, then we may interpolate $P^{[k+1]}$ and recover $Q^{[k+1]}$ using one exact division.

Moreover, we may exploit the assumption that the supports of $P^{[k+1]}$ and $Q^{[k+1]}$ with respect to $x_1, \ldots, x_k$ coincide with the supports of $P^{[k]}$ and $Q^{[k]}$. In the algorithm below, this explains why $m$ evaluation points indeed suffice:

**Algorithm 6.1**

**Input:** $F \in \mathbb{K}[x_1, \ldots, x_n] \setminus \mathbb{K}$ and coprime $A, B \in \mathbb{K}[x_1, x_2] \setminus \mathbb{K}$ with $F^{[2]} = AB$

**Output:** $P, Q \in \mathbb{K}[x_1, \ldots, x_n] \setminus \mathbb{K}$ with $P^{[2]} = A$, $Q^{[2]} = B$, and $F = PQ$, whenever such a factorization exists (we raise an error if this is not the case) and $\pi_{c,2}$ is faithful for this factorization

If $n = 2$, then return $(A, B)$
Compute $P^{[n-1]}, Q^{[n-1]}$ by recursively applying the algorithm to $F^{[n-1]}, A, B$
Permute $P^{[n-1]}$ and $Q^{[n-1]}$ if necessary to ensure that $m := s_{P^{[n-1]}} \leqslant s_{Q^{[n-1]}}$
Compute $F^{\langle n,i\rangle}$, $F^{[n-1,i]}$, and $P^{[n-1,i]}$ for $i = 1, \ldots, m$, using sparse evaluation
Deduce $Q^{[n-1,i]} := F^{[n-1,i]} / P^{[n-1]}$ for $i = 1, \ldots, m$
For $i = 1, \ldots, m$
  Compute $P^{\langle n,i\rangle}, Q^{\langle n,i\rangle}$ with $F^{\langle n,i\rangle} = P^{\langle n,i\rangle} Q^{\langle n,i\rangle}$ using Hensel lifting (Section 4.3)
  Raise an error if this fails
Recover $P$ from $P^{\langle n,1\rangle}, \ldots, P^{\langle n,m\rangle}$ using sparse interpolation
Let $Q := F / P$ and return $(P, Q)$

**Remark 6.4.** The faithfulness assumption implies that $\operatorname{supp} P \subseteq \operatorname{supp} P^{[n-1]} \times \mathbb{N}$. If $d$ is small, then this support bound is reasonably tight. Consequently, we may use sparse interpolation with known supports in order to recover $P$.

THEOREM 6.5. *Let $s := \min(s_P, s_Q)$, $s' := \max(s_P, s_Q)$ $\bar{s} := \max(s', s_F)$, $d := \deg F$, and $\delta := \max(\deg_{x_1} F, \ldots, \deg_{x_n} F)$. Then Algorithm 6.1 runs in time*

$$O(n((\bar{s}/s)\,\mathsf{S}(s') + \delta d\,\mathsf{S}(s) + s\,\mathsf{M}(\delta d) + s\,\mathsf{M}(d)\log d))$$

*and returns the correct result with probability at least*

$$1-O\left(\frac{n s^2 d^2}{N}\right).$$

**Proof.** Assume that we correctly recovered $P^{[n-1]}$ and $Q^{[n-1]}$ and let us investigate the probability that the results $P$ and $Q$ are correct.

Let us first examine the probability that the Hensel lifting method from Section 4.3 works correctly. This is the case whenever the following three conditions are satisfied for $i=1,\ldots,m$:

1. $P^{[n-1,i]}$ and $Q^{[n-1,i]}$ are coprime;
2. We have $\operatorname{cont}_t F^{\langle n,i\rangle}=1$;
3. We have $\deg_t F^{\langle n,i\rangle}=\deg_t F^{[n-1,i]}$.

Consider

$$R := \operatorname{Res}_t\left(\frac{P^{[n-1]}(x_1 t,\ldots,x_{n-1} t)}{t^{\operatorname{val} P^{[n-1]}}},\frac{Q^{[n-1]}(x_1 t,\ldots,x_{n-1} t)}{t^{\operatorname{val} Q^{[n-1]}}}\right) \tag{6.1}$$

with $\deg R\leqslant d^2$. We have $R\neq 0$ since $P^{[n-1]}$ and $Q^{[n-1]}$ are coprime. The first condition fails for a given $i$ if and only if $R$ vanishes under the substitutions $x_1:=\alpha_1^i,\ldots,x_{n-1}:=\alpha_{n-1}^i$. This happens with probability at most $d^2\binom{s}{2}/N$ for some $i$, by Corollary 2.2. Let us next consider

$$R' := \operatorname{Res}_u(F(x_1 t,\ldots,x_{n-1} t,u),F(x_1 t',\ldots,x_{n-1} t',u)),$$

where $x_1,\ldots,x_{n-1},t,t',u$ are formal indeterminates. Since $F$ is content-free, we have $R'\neq 0$. Now $\operatorname{cont}_t F^{\langle n,i\rangle}=1$ whenever $R'$ does not vanish under the substitutions $x_1:=\alpha_1^i,\ldots,$ $x_{n-1}:=\alpha_{n-1}^i$. The second condition therefore fails for some $i$ with probability at most $4d^2\binom{s}{2}/N$, by Corollary 2.2 and using the fact that $\deg R'\leqslant 4d^2$. Finally, with probability at least $1-d\binom{s}{2}/N$, we have

$$\deg_t F^{\langle n,i\rangle} = \deg_t F^{[n-1,i]} = \deg_t F(x_1 t,\ldots,x_{n-1} t,x_n)$$

for $i=1,\ldots,m$, by Corollary 2.2.

Let us now return to the correctness of the overall algorithm. Concerning the top-level call, we have the following:

- Obtaining $Q^{[n-1,i]}$ from the evaluations $F^{[n-1,i]}$, $P^{[n-1,i]}$ for $i=1,\ldots,m$ by exact univariate polynomial division is possible as long as $\deg F^{[n-1,i]}=\deg F^{[n-1]}$. This condition fails with probability at most $s^2 d/N$ by Corollary 2.2.
- We have shown above that the Hensel lifting strategy from Section 4.3 can be applied with probability at least $1-O(d^2 s^2/N)$. By adapting the proof of Proposition 4.3 such that we can apply Corollary 2.2, the Hensel lifting itself also succeeds with probability at least $1-O(d^2 s^2/N)$, for $i=1,\ldots,m$.

Altogether, we correctly determine $P$ and $Q$ with probability at least $1-O(s^2 d^2/N)$. Since there are $O(n)$ recursive levels and since the sizes of the supports of $F^{[k]}$, $P^{[k]}$, and $Q^{[k]}$ are all smaller than those of $F$, $P$, and $Q$, the probabilistic correctness claim follows.

As to the complexity bound, let us now study the cost when not counting the recursive calls:

- The computation of the $F^{\langle n,i\rangle}$, $F^{[n-1,i]}$, and $P^{[n-1,i]}$, using sparse evaluation at geometric sequences requires $O((\bar{s}/s+d\deg_{x_n} F)\,\mathsf{S}(s))=O(\mathsf{S}(\bar{s})+\delta d\,\mathsf{S}(s))$ operations in $\mathbb{K}$.

- The evaluations of $Q^{[k-1]}$ are obtained by exact univariate polynomial division and require $O(s\,\mathsf{M}(d))$ operations in total.
- Lifting the factorization for all of the evaluations requires $O(s\,(\mathsf{M}(\delta d)+\mathsf{M}(d)\log d))$ further operations, by Proposition 4.3.
- We obtain $P=P_0+\cdots+P_\delta x_n^\delta$ from $P^{\langle n,1\rangle},\ldots,P^{\langle n,m\rangle}$ using sparse interpolation of the coefficients $P_0,\ldots,P_\delta$ in time $O(\mathsf{S}(s_{P_0})+\cdots+\mathsf{S}(s_{P_\delta}))=O(\mathsf{S}(s)+s\,\delta)=O(\delta\,\mathsf{S}(s))$.
- The recovery of $Q$ through sparse polynomial division takes $O((\bar{s}/s')\,\mathsf{S}(s'))$ further operations, as explained in Section 2.4.

We conclude that the cost is $O((\bar{s}/s')\,\mathsf{S}(s')+\delta d\,\mathsf{S}(s)+s\,\mathsf{M}(\delta d)+s\,\mathsf{M}(d)\log d)$, when not counting the recursive calls. Since there are $O(n)$ recursive levels and since the sizes of the supports of $F^{[k]}$, $P^{[k]}$, and $Q^{[k]}$ are all smaller than those of $F$, $P$, and $Q$, the complexity bound follows. □

**Example 6.6.** Let $F=P\,Q$, where $P,Q\in\mathbb{K}[x_1,\ldots,x_n]$ are coprime. Now consider the polynomial $\tilde{F}(x_1,\ldots,x_{2n})=F(x_1x_{n+1},x_2x_{n+2},\ldots,x_nx_{2n})$. When factoring $F$ using Algorithm 6.1, the last $n$ lifting steps with respect to $x_{n+1},\ldots,x_{2n}$ all operate on a projection $F^{[k]}$ of size $s_F$. This is an example when the overhead $n$ in the complexity bound is sharp. However, the support of $F$ is very special, since it is contained in an affine subspace of dimension $\leqslant n$. It is easy to detect this situation and directly obtain the exponents in the last $n$ variables as affine combinations of the known exponents in the first $n$ variables.

More generally, it may happen that the fibers of the support under the projection with respect to the last $k$ variables are all small. It would be interesting to develop specific optimizations for this situation, e.g. directly apply sparse interpolation on the fibers.

## 6.3. Simultaneous lifting of multiple factors

The lifting algorithm generalizes in a straightforward way to the case when $F$ is a product of more than two factors. This time, we wish to recover a factorization $F=\lambda\,P_1^{\nu_1}\cdots P_\ell^{\nu_n}$ from $A_1:=P_1^{[2]},\ldots,A_\ell:=P_\ell^{[2]}$, assuming that $A_1,\ldots,A_\ell$ (hence $P_1,\ldots,P_\ell$) are pairwise coprime.

**Algorithm 6.2**

**Input:** $F\in\mathbb{K}[x_1,\ldots,x_n]\setminus\mathbb{K}$, irreducible and pairwise coprime $A_1,\ldots,A_\ell\in\mathbb{K}[x_1,x_2]\setminus\mathbb{K}$, $\ell\geqslant 2$, $\lambda\in\mathbb{K}^{\neq}$, and $\nu_1,\ldots,\nu_\ell\in\mathbb{N}$, such that $F^{[2]}=\lambda\,A_1^{\nu_1}\cdots A_\ell^{\nu_\ell}$

**Output:** irreducible and pairwise coprime $P_1,\ldots,P_\ell\in\mathbb{K}[x_1,\ldots,x_n]\setminus\mathbb{K}$ with $F=\lambda\,P_1^{\nu_1}\cdots P_\ell^{\nu_\ell}$ and $P_1^{[2]}=A_1,\ldots,P_\ell^{[2]}=A_\ell$, whenever such a factorization exists (we raise an error if this is not the case) and $\pi_{c,2}$ is faithful for this factorization

If $n=2$, then return $(A_1,\ldots,A_\ell)$
Recursively apply the algorithm to $F^{[n-1]},A_1,\ldots,A_\ell$ to compute $P_1^{[n-1]},\ldots,P_\ell^{[n-1]}$
Let $m:=\max(s_{P_1^{[n-1]}},\ldots,s_{P_\ell^{[n-1]}})$
Compute $F^{\langle n,i\rangle}$, $P_1^{[n-1,i]},\ldots,P_\ell^{[n-1,i]}$ for $i=1,\ldots,m$, using sparse evaluation
Deduce $(P_1^{[n-1,i]})^{\nu_1},\ldots,(P_\ell^{[n-1,i]})^{\nu_\ell}$ for $i=1,\ldots,m$
For $i=1,\ldots,m$
    Compute $P_1^{\langle n,i\rangle},\ldots,P_\ell^{\langle n,i\rangle}$ with $F^{\langle n,i\rangle}=\lambda\,(P_1^{\langle n,i\rangle})^{\nu_1}\cdots(P_\ell^{\langle n,i\rangle})^{\nu_\ell}$ using Hensel lifting
    Raise an error if this fails
Recover $P_j$ from the $P_j^{\langle n,i\rangle}$ using sparse interpolation, for $j=1,\ldots,\ell$
Return $(P_1,\ldots,P_\ell)$

**Remark 6.7.** More precisely, we compute $(P_j^{[n-1,i]})^{\nu_j}$ using binary powering and $P_j^{[n,i]}$ from $(P_j^{[n,i]})^{\nu_j}$ using dense bivariate $\nu_j$-th root extraction, for $j=1,\ldots,\ell$. As an optimization, we may first sort the factors of $F^{[n-1]}$ such that $s_{P_1^{[n-1]}} \leqslant s_{P_2^{[n-1]}} \leqslant \cdots \leqslant s_{P_\ell^{[n-1]}}$ and then compute $(P_\ell^{[n-1,i]})^{\nu_\ell} := F^{[n-1,i]} / ((P_1^{[n-1,i]})^{\nu_1} \cdots (P_{\ell-1}^{[n-1,i]})^{\nu_{\ell-1}})$ using exact division.

THEOREM 6.8. *Let $s := \max(s_{P_1},\ldots,s_{P_\ell})$, $\bar{s} := \max(s, s_F)$, $\delta := \max(\deg_{x_1} F,\ldots,\deg_{x_n} F)$, and $d := \deg F$. Then Algorithm 6.2 runs in time*

$$O(n((\bar{s}/s + \delta d)\,\mathsf{S}(s) + s\,\mathsf{M}(\delta d)\log \ell + s\,\mathsf{M}(d)\log d))$$

*and returns the correct result with probability at least*

$$1 - O\left(\frac{n\,\ell s^2 d^2}{N}\right).$$

**Proof.** The proof is similar to that of Algorithm 6.1. Assume now that we correctly recovered $P_1^{[n-1]},\ldots,P_\ell^{[n-1]}$ and let us investigate the probability that the results $P_1,\ldots,P_\ell$ are correct. To apply the Hensel lifting method from Section 4.3, the following three conditions must be satisfied for $i=1,\ldots,m$:

1. $P_1^{[n-1,i]},\ldots,P_\ell^{[n-1,i]}$ are pairwise coprime;
2. We have $\operatorname{cont}_t F^{\langle n,i\rangle} = 1$;
3. We have $\deg_t F^{\langle n,i\rangle} = \deg_t F^{[n-1,i]}$.

The analysis is similar as in the proof of Theorem 6.5, except that (6.1) now becomes

$$R := \prod_{1 \leqslant i < j \leqslant \ell} \operatorname{Res}_t\left(\frac{P_i^{[n-1]}(x_1 t,\ldots,x_{n-1} t)}{t^{\operatorname{val} P_i^{[n-1]}}}, \frac{P_j^{[n-1]}(x_1 t,\ldots,x_{n-1} t)}{t^{\operatorname{val} P_j^{[n-1]}}}\right)$$

and the probability that $R$ vanishes for one of the $m$ substitutions is now bounded by $O(\ell s^2 d^2/N)$, since $\deg R \leqslant 2 d^2 \ell$. Using the fact there are $\leqslant n$ recursive levels, this completes the probabilistic correctness proof.

For the cost analysis, we again start by analyzing the cost of the algorithm without the recursive calls:

- The evaluation of the factors $P_1^{[n-1]},\ldots,P_\ell^{[n-1]}$ on the geometric sequence can be done in time $O((m/s_{P_1}+\delta)\,\mathsf{S}(s_{P_1}) + \cdots + (m/s_{P_\ell}+\delta)\,\mathsf{S}(s_{P_\ell})) = O(\delta\,\ell\,\mathsf{S}(s)) = O(\delta d\,\mathsf{S}(s))$.
- The computation of the $F^{\langle n,i\rangle}$ and $F^{[n-1,i]}$ takes $O((s_F/s+\delta d)\,\mathsf{S}(s))$ operations.
- The powers $(P_1^{[n-1,i]})^{\nu_1},\ldots,(P_\ell^{[n-1,i]})^{\nu_\ell}$ can be obtained in time $O(s(\mathsf{M}(\nu_1 d_{P_1}) + \cdots + \mathsf{M}(\nu_\ell d_{P_\ell}))) = O(s\,\mathsf{M}(d))$, using binary powering.
- The Hensel lifting is done using Proposition 4.4 instead of Proposition 4.3, which contributes $O(s(\mathsf{M}(d\delta)\log \ell + \ell\,\mathsf{M}(\delta)\log \delta))$ to the cost.
- The bivariate root extractions take $O(s(\mathsf{M}(\delta \nu_1 d_{P_1}) + \cdots + \mathsf{M}(\delta \nu_\ell d_{P_\ell}))) = O(s\,\mathsf{M}(\delta d))$ operations, by Proposition 4.2.

- The recovery of the $P_j$ from the $P_j^{\langle n,i \rangle}$ using sparse interpolation requires $O(\mathsf{S}(s_{P_1})+\cdots+\mathsf{S}(s_{P_\ell}))=O(d\,\mathsf{S}(s))$ operations.

Summing these costs and multiplying with the recursive depths $O(n)$ yields the desired complexity bound. □

**Example 6.9.** There are cases when the recursive application of Algorithm 6.1 may lead to intermediate expression swell. For instance, for $i=1,2$ and $k\in\mathbb{N}$, consider the polynomials

$$\begin{aligned} P_i &:= (x_i+y_i)^k-(u_i+v_i)^k \\ &= (x_i+y_i-u_i-v_i)\,Q_i \\ Q_i &:= \sum_{0\leqslant j<k} (x_i+y_i+u_i+v_i)^j \end{aligned}$$

and note that $Q_i$ has $\Theta(k^3)$ terms, whereas $P_i$ has only $O(k)$ terms. Now consider

$$F = P_1 P_2.$$

Then a direct factorization of $F$ into irreducibles can be done in time $\tilde{O}(k^3)$, whereas recursively factoring out $(x_1+y_1-u_1-v_1)$ and then $(x_2+y_2-u_2-v_2)$ may lead us to consider intermediate expressions like $Q_1 Q_2$ of size $\Theta(k^6)$.

**Remark 6.10.** The ideas behind Algorithm 6.2 can readily be adapted to $p$-adic lifting, in order to factor a polynomial $F\in\mathbb{Q}[x_1,\ldots,x_n]$ with rational coefficients. In that case, we start with a factorization $\bar{F}=\bar{c}\,\bar{P}_1^{\nu_1}\cdots\bar{P}_\ell^{\nu_\ell}$ of $\bar{F}:=F \bmod p\in\mathbb{F}_p[x_1,\ldots,x_n]$ for some sufficiently large prime $p$. The goal is to lift this into a factorization of $F \bmod p^\kappa$ for some sufficiently large $\kappa$ and then to recover a factorization of $F$ using rational number reconstruction [31, Section 5.10]. The relevant analogues of $A^{\langle n,i\rangle}(t,u)$ and $A^{[n-1,i]}(t)$ are $A^{\langle i\rangle}(t):=P(\alpha_1^i\,t,\ldots,\alpha_n^i\,t)$ and $A^{[i]}(t):=A^{\langle i\rangle}(t) \bmod p$, where the prime number $p$ plays a similar role as the formal indeterminate $u$. The bivariate Hensel lifting in $K[[u]][t]$ is replaced by traditional univariate Hensel lifting in $\mathbb{Z}_p[t]$.

## 6.4. Irreducible factorization

In lucky cases, given an irreducible factorization $F=\lambda\,P_1^{\nu_1}\cdots P_\ell^{\nu_\ell}$, random choices of $c_1,\ldots,c_\ell$ give rise with high probability to an irreducible factorization $F^{[2]}=\lambda\,(P_1^{[2]})^{\nu_1}\cdots(P_\ell^{[2]})^{\nu_\ell}$. For such $F$, we may use the following algorithm to compute its irreducible factorization:

**Algorithm 6.3**
**Input:** a content-free polynomial $F\in\mathbb{K}[x_1,\ldots,x_n]\setminus\mathbb{K}$
**Output:** an irreducible factorization $F=\lambda\,P_1^{\nu_1}\cdots P_\ell^{\nu_\ell}$ or an error

If $n\leqslant 1$, then return the result of a univariate irreducible factorization
Let $c_1,\ldots,c_n$ be random elements of $\mathbb{K}^{\neq}$
Compute an irreducible factorization $F^{[2]}=\lambda\,A_1^{\nu_1}\cdots A_\ell^{\nu_\ell}$ of $F^{[2]}$
Apply Algorithm 6.2 to $F$ and this factorization

THEOREM 6.11. *Let* $s := \max(s_{P_1}, \ldots, s_{P_\ell})$, $\bar{s} := \max(s, s_F)$, $\delta := \max(\deg_{x_1} F, \ldots, \deg_{x_n} F)$, *and* $d := \deg F$. *Then Algorithm 6.3 runs in time*

$$O(n((\bar{s}/s + \delta d)\,\mathsf{S}(s) + s\,\mathsf{M}(\delta d)\log \ell + s\,\mathsf{M}(d)\log d)) + \tilde{O}(\delta^3) + \mathsf{F}(\delta)$$

*and returns the correct result (when no error is raised) with probability at least*

$$1 - O\!\left(\frac{n\,\ell s\,(s+n)\,d^2}{N}\right).$$

**Proof.** This follows from Theorems 6.8, 4.7, and Lemma 6.2. □

**Remark 6.12.** We have seen in Section 5.4 that the size of the content of a multivariate polynomial can be much larger than the size of the polynomial itself, in pathological cases. In such cases, we may wish to avoid content-free factorization as a first step of a general algorithm for irreducible factorization. This can be achieved by adapting Algorithm 6.2 so as to factor out the content $C$ in $x_n$ at the top-level and performing the recursive call and bivariate lifting to $F/C$ instead of $F$. Incorporating the content-free factorization directly into the main algorithm in this way yields a similar complexity bound as in Theorem 6.11, but with $\mathsf{F}(d)$ instead of $\mathsf{F}(\delta)$.

**Remark 6.13.** An alternative approach for reducing to the content-free case, which also works for the algorithms from Section 7 below, is to tag $F$ according to a regularizing weight $w$. This has the advantage of "exposing" the entire factorization of $F$ with respect to the tagging variable $t$. Of course, this requires to replace the degree $d = d_F$ in the complexity bound by $e = \mathrm{ec}_w F \leqslant d^2$.

**Remark 6.14.** The problem from Remark 6.3 can be partially remedied by amending the bivariate lifting step: instead of raising an error when $F^{\langle n,i\rangle} \neq \lambda\,(P_1^{\langle n,i\rangle})^{\nu_1}\cdots(P_\ell^{\langle n,i\rangle})^{\nu_\ell}$, we may continue the Hensel lifting to a higher precision (the $P_j^{\langle n,i\rangle}$ are now power series in $\mathbb{K}[t][[u]]$) and determine which factors need to be combined in order to obtain a factorization of $F^{\langle n,i\rangle}$ (using the techniques from [74]). Doing this for all $i$ leads to a finest partition $J_1 \amalg \cdots \amalg J_{\ell'}$ of $\{1,\ldots,\ell\}$ such that $\Pi_k^{\langle n,i\rangle} := \prod_{j\in J_k} P_j^{\langle n,i\rangle}$ is a polynomial factor of $F^{\langle n,i\rangle}$ for $i = 1, \ldots, m$ and $k = 1, \ldots, \ell'$. We next apply the sparse interpolation to the polynomials $\Pi_k^{\langle n,i\rangle}$ instead of the series $P_j^{\langle n,i\rangle}$. This may require more than $m$ interpolation points, so we also double $m$ until the interpolation step succeeds. Combined with Remark 6.12, this yields to an algorithm of complexity

$$O(n((\bar{s}/s + \delta d)\,\mathsf{S}(s) + s\,\tilde{O}(d\,\delta^2))) + \mathsf{F}(d),$$

where $s$ is the maximal size of an irreducible factor of $F^{[k]}$ for $k \in \{1,\ldots,n\}$.

Unfortunately, we have not yet been able to perform a clean probabilistic analysis for this method. Indeed, we still need to prove that the algorithm terminates in reasonable expected time for irreducible polynomials. Now consider the example

$$F \;=\; x_1^2 + x_2^2 - x_3^2, \tag{6.2}$$

for which we obtain

$$F^{\langle 3,i\rangle}(t,u) \;=\; (\alpha_1^{2i} + \alpha_2^{2i})\,t^2 - u^2.$$

Whenever $\mathbb{K}=\mathbb{F}_p$ with $p\geqslant 3$ and $\alpha_1^{2i}+\alpha_2^{2i}$ is a square in $\mathbb{K}$, the projected polynomial $F^{\langle 3,i\rangle}$ is not irreducible. In principle, this only happens with probability $1/2$, so this is unlikely to happen for all $m$ values of $i=1,\ldots,m$. However, we were unable so far to prove a general bound. We also note that this probabilistic argument is different from traditional Hilbert–Bertini style arguments [100, Section 6.1]; see also [74] and [75, Chapître 6].

**Remark 6.15.** Another way to turn Algorithm 6.3 into a method that always succeeds with high probability would be to apply random monomial transformation to $F$. For instance, after the change of variables $x_1=y_1y_2y_3$, $x_2=y_2y_3$, $x_3=y_3$, we have

$$F = (y_1^2y_2^2+y_2^2-1)\,y_3^2,$$

for the polynomial $F$ from example (6.2). With high probability, Algorithm 6.3 returns the correct irreducible factorization of $F$ after this rewriting.

## 7. FACTORING USING PROJECTIVE HENSEL LIFTING

The iterative approach from Sections 3.1 and 6 has the disadvantage of introducing a dependence on the dimension $n$ in the complexity bounds. In the case of gcd computations, the idea of regularizing weights allowed for the more direct approach in Section 3.2.

Unfortunately, direct adaptations of this approach to factorization only work in very special cases, because it is unclear how to recombine the factors found at different evaluation points. One notable exception is when $F\in\mathbb{K}[x_1,\ldots,x_n]\setminus\mathbb{K}$ factors, say, into two irreducible factors of degrees one and two in one of the variables; in that case, we can recombine based on degree information.

In this section, we will develop another "direct" approach for the factorization of $F\in\mathbb{K}[x_1,\ldots,x_n]\setminus\mathbb{K}$ that avoids iteration on the dimension $n$. Except for precomputations in the algorithm from Section 7.7, the algorithms in this section do not rely on regularizing weights, but exploits more subtle properties of the Newton polytope of $F$. We will present three versions of our approach in order to deal with Newton polytopes of increasing complexity. For simplicity, we only present them in the case when $F$ is the product of two factors. As we shall see in Section 7.6, even the last and most elaborate version of our approach is not fully general. We will conclude with a theoretical variant that is less efficient in practice, but which has the advantage of providing a full algorithm for the computation of irreducible factorizations.

### 7.1. A favorable special case

Assume that the unique factorization of $F$ contains exactly two factors

$$F(x_1,\ldots,x_n) = P(x_1,\ldots,x_n)\,Q(x_1,\ldots,x_n), \qquad (7.1)$$

where $P,Q\in\mathbb{K}[x_1,\ldots,x_n]$ are irreducible and distinct and both depend on the "main" variable, say, $x_1$. In particular, this implies that $F$ is content-free and square-free. Assume also that the following conditions hold:

**H1.** $\deg_{x_1}F(x_1,0,\ldots,0)=\deg_{x_1}F$.

**H2.** $P(x_1,0,\ldots,0)$ and $Q(x_1,0,\ldots,0)$ are coprime.

Note that **H1** allows us to normalize the factorization

$$F(x_1,0,\ldots,0) \;=\; P(x_1,0,\ldots,0)\,Q(x_1,0,\ldots,0)$$

by requiring that $P(x_1,0,\ldots,0)$ is monic. From now on we will always assume that we have done this.

Under these assumptions, we claim that $P$ and $Q$ can be computed from $P(x_1,0,\ldots,0)$ and $Q(x_1,0,\ldots,0)$ using Hensel lifting and sparse interpolation. In order to see this, let $\alpha=(1,\alpha_2,\ldots,\alpha_n)\in(\mathbb{K}^{\neq})^n$ be an admissible ratio or an FFT ratio. For each $i\in\mathbb{N}$, we define bivariate polynomials $F^{[i]},P^{[i]},Q^{[i]}\in\mathbb{K}[x_1,t]$ by

$$\begin{aligned} F^{[i]} &:= F(x_1,\alpha_2^i t,\ldots,\alpha_n^i t)\\ P^{[i]} &:= P(x_1,\alpha_2^i t,\ldots,\alpha_n^i t)\\ Q^{[i]} &:= Q(x_1,\alpha_2^i t,\ldots,\alpha_n^i t). \end{aligned}$$

Then $P^{[i]}(x_1,0)=P(x_1,0,\ldots,0)$ and $Q^{[i]}(x_1,0)=Q(x_1,0,\ldots,0)$ are coprime and we have $F^{[i]}=P^{[i]}Q^{[i]}$. This allows us to compute $P^{[i]}$ and $Q^{[i]}$ from $F^{[i]}$, $P^{[i]}(x_1,0)$, and $Q^{[i]}(x_1,0)$ using Hensel lifting. For sufficiently large $m$, with $m=O(\min(s_P,s_Q))$, we may then use sparse interpolation to recover $P$ from $P^{[1]},\ldots,P^{[m]}$ or $Q$ from $Q^{[1]},\ldots,Q^{[m]}$. This leads to the following algorithm:

**Algorithm 7.1**

**Input:** $F\in\mathbb{K}[x_1,\ldots,x_n]$ such that there exists a factorization (7.1) that satisfies **H1** and **H2**; we assume that $P(x_1,0,\ldots,0)$ and $Q(x_1,0,\ldots,0)$ are given

**Output:** $P,Q\in\mathbb{K}[x_1,\ldots,x_n]$

For $m=1,2,4,8,\ldots$ do
- Compute $F^{[i]}$ for $i=\lfloor m/2\rfloor+1,\ldots,m$
- Compute $P^{[i]},Q^{[i]}$ for $i=\lfloor m/2\rfloor+1,\ldots,m$,
  using bivariate Hensel lifting for $F^{[i]}$, $P^{[1]}(x_1,0)$, $Q^{[1]}(x_1,0)$
- If $P^{[1]},\ldots,P^{[m]}$ yield $P$ through sparse interpolation, then return $(P,F/P)$
- If $Q^{[1]},\ldots,Q^{[m]}$ yield $Q$ through sparse interpolation, then return $(F/Q,Q)$

THEOREM 7.1. *Let* $s:=\min(s_P,s_Q)$, $s':=\max(s_P,s_Q)$, $\bar{s}:=\max(s',s_F)$, $\delta:=\deg_{x_1}F$, *and* $d:=\deg F$. *Then Algorithm 7.1 is correct with probability at least* $1-d\binom{s'}{2}/N$ *and runs in time*

$$O((\bar{s}/s')\,\mathsf{S}(s')+\delta d\,\mathsf{S}(s)+s\,\mathsf{M}(\delta d)+s\,\mathsf{M}(d)\log d).$$

**Proof.** The correctness is clear. Let us analyze the cost of the main steps of the algorithm:

- The computation of the $F^{[i]}$ takes $O((s_F/s+\delta d)\,\mathsf{S}(s))$ operations.
- The Hensel lifting has a total cost of $O(s\,(\mathsf{M}(\delta d)+\mathsf{M}(d)\log d))$.
- The sparse interpolation of $P$ (or $Q$) takes $O(\delta d\,\mathsf{S}(s))$ operations.
- The sparse polynomial division to recover $Q$ (or $P$) requires $O((\bar{s}/s')\,\mathsf{S}(s'))$ operations.

Adding up these costs, the complexity bound follows. The algorithm is correct as long as the final division succeeds, which is the case with probability at least $1-d\binom{s'}{2}/N$. □

## 7.2. Single slope bivariate Hensel lifting

The assumptions **H1** and **H2** are fairly strong: they are never satisfied both if $F$ is a homogeneous polynomial. Now the assumptions **H1** and **H2** essentially allow us to perform the bivariate Hensel lifting using the method from Section 4.3. The problem that we face in order to extend the approach from Algorithm 7.1 is that random shifts $y \mapsto \sigma + y$ are not allowed for the bivariate Hensel lifting from Section 4.3: we only have an efficient algorithm to compute $P(x,0)$ and $Q(x,0)$ are given, not $P(x,\sigma)$ and $Q(x,\sigma)$.

Instead, we need a way to generalize the algorithm from Section 4.3 to the case when the Newton polygon of $F$ is non-trivial. In this subsection, we start with the simplest "single slope" case. Since this material is a supplement to Section 4.3, we adopt the notation from there.

Assume that $F \in \mathbb{K}[x,y]$ has a non trivial factorization $F = PQ$ with $P,Q \in \mathbb{K}[x,y] \setminus \mathbb{K}$. We define the *Newton polygon* of $F$ to be the "lower border" of its Newton polytope:

$$\mathcal{N} \;:=\; \operatorname{npol} P \;:=\; \partial_{\text{low}} \operatorname{hull} P \;:=\; \operatorname{hull} P \setminus (\operatorname{hull} P + \{0\} \times \mathbb{R}^{>}).$$

It is the union of a finite number of edges $E_1,\ldots,E_\ell$ between $(i_0,j_0),\ldots,(i_\ell,j_\ell) \in \operatorname{supp} P$ with $i_0 = \operatorname{val}_x P$, $i_\ell = \deg_x P$, and $i_0 < \cdots < i_\ell$. For each edge $E_k$, there is a corresponding weight $w_k := ((i_{k-1} - i_k) / (j_k - j_{k-1}), 1)$ such that $E_k = \operatorname{hull} \operatorname{lp}_{w_k} P$.

Assume that $\mathcal{N}$ has a single edge, let $w := w_1$, and let $p \in \mathbb{Z}$ and $q \in \mathbb{N}^{>}$ be coprime such that $p / q = (i_0 - i_1) / (j_1 - j_0)$. Now consider

$$\begin{aligned}
\tilde{F}(x,t) &:= F(x t^p, t^q)\, t^{-q \operatorname{val}_w F}\\
\tilde{P}(x,t) &:= P(x t^p, t^q)\, t^{-q \operatorname{val}_w P}\\
\tilde{Q}(x,t) &:= Q(x t^p, t^q)\, t^{-q \operatorname{val}_w Q}.
\end{aligned}$$

Then $\tilde{F}, \tilde{P}, \tilde{Q} \in \mathbb{K}[x,t]$ and $\tilde{F} = \tilde{P} \tilde{Q}$. Moreover, $\tilde{F}(x,0) = (\operatorname{tp}_w F)(x t^p, t^q)\, t^{-q \operatorname{val}_w F}$, and we have similar expressions for $\tilde{P}(x,0)$ and $\tilde{Q}(x,0)$. If $\operatorname{tp}_w P$ and $\operatorname{tp}_w Q$ are known and if their transforms $\tilde{P}(x,0)$ and $\tilde{Q}(x,0)$ are coprime, then this enables us to compute $P$ and $Q$ by applying the Hensel lifting method from Section 4.3 to $\tilde{F}$, $\tilde{P}(x,0)$, and $\tilde{Q}(x,0)$.

## 7.3. Single slope factorization

Let us now return to the factorization problem from Section 7.1 and let us show how to generalize Algorithm 7.1 using the material from the previous subsection.

Let $S := \{(e_1, e_2 + \cdots + e_n) : e \in \operatorname{supp} F\}$ and let $\mathcal{N}$ be the lower boundary of its convex hull. With high probability, we have $\operatorname{supp} F^{[i]} = S$ and $\operatorname{npol} F^{[i]} = \mathcal{N}$, for any given $i$. The last edge of $\mathcal{N}$ joins the points $(a',a)$ and $(b',b)$ for $a,b,a',b' \in \mathbb{N}$ with $a' < b' = \deg_{x_1} F$. Let $w = (w_1,\ldots,w_n)$ with $w_1 = p / q = (a-b) / (b'-a')$ and $w_2 = \cdots = w_n = 1$, where $p \in \mathbb{Z}$, $q \in \mathbb{N}^{>}$, and $\gcd(p,q) = 1$. Now consider the assumptions

**H1'.** $\deg_{x_1} \operatorname{tp}_w F = \deg_{x_1} F$.

**H2'.** $\operatorname{tp}_w P$ and $\operatorname{tp}_w Q$ are coprime.

The hypotheses **H1** and **H2** correspond to the special case when $a = b = 0$. If $\operatorname{tp}_w P$ and $\operatorname{tp}_w Q$ are known (e.g. through the recursive factorization of $\operatorname{tp}_w F$ if $F$ is not $w$-homogeneous), then the algorithm from Section 7.2 allows us to Hensel lift the factorization $\operatorname{tp}_w F = (\operatorname{tp}_w P)(\operatorname{tp}_w Q)$ into a factorization $F = PQ$. This leads to the following generalization of Algorithm 7.1:

**Algorithm 7.2**

**Input:** $F \in \mathbb{K}[x_1, \ldots, x_n]$ such that there exists a factorization (7.1) that satisfies **H1'** and **H2'**, together with $\operatorname{tp}_w P$ and $\operatorname{tp}_w Q$ such that $\operatorname{tp}_w F = (\operatorname{tp}_w P)(\operatorname{tp}_w Q)$

**Output:** $P, Q \in \mathbb{K}[x_1, \ldots, x_n]$

For $m = 1, 2, 4, 8, \ldots$ do
  Compute $F^{[i]}$, $(\operatorname{tp}_w P)^{[i]}$, $(\operatorname{tp}_w Q)^{[i]}$ for $i = \lfloor m/2 \rfloor + 1, \ldots, m$
  Compute $P^{[i]}$, $Q^{[i]}$ for $i = \lfloor m/2 \rfloor + 1, \ldots, m$,
    using bivariate Hensel lifting from Section 7.2 for $F^{[i]}$, $(\operatorname{tp}_w P)^{[i]}$, $(\operatorname{tp}_w Q)^{[i]}$
  If $P^{[1]}, \ldots, P^{[m]}$ yield $P$ through sparse interpolation, then return $(P, F/P)$
  If $Q^{[1]}, \ldots, Q^{[m]}$ yield $Q$ through sparse interpolation, then return $(F/Q, Q)$

THEOREM 7.2. *Let* $s := \min(s_P, s_Q)$, $s' := \max(s_P, s_Q)$, $\bar{s} := \max(s', s_F)$, $\delta := \deg_{x_1} F$, *and* $d := \deg F$. *Then Algorithm 7.2 is correct with probability at least* $1 - d\binom{s'}{2}/N$ *and runs in time*

$$O((\bar{s}/s')\,\mathsf{S}(s') + \delta d\,\mathsf{S}(s) + s\,\mathsf{M}(\delta d) + s\,\mathsf{M}(d)\log d).$$

**Proof.** The proof is similar to the proof of Theorem 7.1. The main change is that the generalized algorithm also requires the evaluations of $\operatorname{tp}_w P$ and $\operatorname{tp}_w Q$ on our sequence. This takes $O((s_P/s + \delta d)\,\mathsf{S}(s) + (s_Q/s + \delta d)\,\mathsf{S}(s)) = O((\bar{s}/s + \delta d)\,\mathsf{S}(s))$ additional operations, which is absorbed by the complexity bound. □

Note that the assumptions **H1'** and **H2'** are actually slightly more liberal than the statement that the Newton polygon should have a single edge. For this reason, Algorithm 7.2 is very likely to apply as soon as there exists a variable $x_i$ for which $\deg_{x_i} F$ is small (of course, we may then assume $i = 1$ modulo a permutation of variables). This in particular the case when $\deg_{x_i} F = 2$, unless $\operatorname{tp}_w F$ is not square-free (one may need to replace $x_i \mapsto x_i^{-1}$ and multiply $F$ with $x_i^2$). Note that a single "good" variable $x_i$ suffices.

**Remark 7.3.** Both Algorithms 6.2 and 7.2 involve recursive factorizations of polynomials in less variables. However, the supports of the recursive calls are obtained through projection for Algorithm 6.2 and through restriction for Algorithm 7.2. The second case is often more favorable in the sense that the recursive supports decrease faster in size.

**Remark 7.4.** Consider the case when $P = 1 + x_1 x_2$ and $Q = 1 + 2 x_1 x_2$. Then $\operatorname{tp}_w F$ coincides with $F$, so the requested factorization of $\operatorname{tp}_w F$ is not simpler than the desired factorization of $F$. This is due to the fact that the exponents of $F$ all lie in a linear subspace of $\mathbb{Z}^n$. In particular, the degree in $x_1$ of any term of $F$ can be determined as a function of the degrees of the other variables. In particular, we may recover a factorization of $F$ from a factorization of $F(1, x_2, \ldots, x_n)$. Using a straightforward induction on $n$, this allows us to ensure that $\operatorname{tp}_w F$ has strictly less terms than $F$.

## 7.4. Multiple slope bivariate Hensel lifting

As soon as $\deg_{x_1} F \geqslant 4$, it can happen that every factor of $F$ has a Newton polygon with at least two edges. In order to deal with this situation, the algorithm from Section 7.2 may be further generalized to accommodate Newton polygons with multiple slopes. In order to explain how, we again adopt the notation from Sections 4.3 and 7.2.

So assume that $P(x,0) \neq 0$ and $P(0,y) \neq 0$, that $\mathcal{N}$ has an arbitrary number of edges, and that $\mathrm{tp}_{w_k} P$ and $\mathrm{tp}_{w_k} Q$ are known for each edge $E_k$. Assume also that $\xi(\mathrm{tp}_{w_k} P)$ and $\xi(\mathrm{tp}_{w_k} Q)$ are coprime for $k = 1, \ldots, \ell$, where $\xi \colon \mathbb{K}[x,y] \setminus \mathbb{K} \to \mathbb{K}[x,y] \setminus \mathbb{K}$ is the normalization mapping with $\xi(P) := P / (x^{\mathrm{val}_x P} y^{\mathrm{val}_y P})$. We will say that $(\mathrm{tp}_{w_k} P)\,(\mathrm{tp}_{w_k} Q)$ is a *coprime edge factorization* of $\mathrm{tp}_{w_k} F$. The aim of this subsection is to lift these factorizations efficiently into the global factorization $F = P\,Q$.

It is well known that a variant of the Newton polygon method can be used in order to factor $F$ over the ring $\mathbb{K}((y))$ of Laurent series. An efficient algorithm for this task was described in [55]. One important subtask is distinct slope factorization: each edge $E_k$ gives rise to a unique monic factor $A_k \in \mathbb{K}((y))[x]$ of $F$ such that $\mathrm{tp}_{w_k} F = c x^a y^b \mathrm{tp}_{w_k} A_k$ for $c \in \mathbb{K}^{\neq}$, $a, b \in \mathbb{N}$, and the natural generalization of the notation $\mathrm{tp}_{w_k}$ to $\mathbb{K}((y))[x]$. The methods from [55] allow for the efficient computation of the $A_k$: it takes time $O(\mathsf{M}(d_x d_y) \log d_x)$ to compute $A_k y^{-\mathrm{val}_y A_k} \in \mathbb{K}[[y]][x]$ at order $d_y$ for $k = 1, \ldots, \ell$.

Now for each $k \in \{1, \ldots, \ell\}$, the known factor $\mathrm{tp}_{w_k} P$ of $\mathrm{tp}_{w_k} F$ induces a factor of $\mathrm{tp}_{w_k} A_k$ that can be Hensel lifted into the factor $B_k := \gcd(P, A_k)$ of $A_k$, by adapting the techniques from Section 4.3 to Laurent series coefficients. For some $C \in \mathbb{K}((y))$, we then have $P = C\,B_1 \cdots B_\ell$. We may determine $C$ in a similar way as in Section 4.3. The other factor $Q := F / P$ can be obtained using one bivariate division. The total cost of this method to compute $P$ and $Q$ is bounded by $O(\mathsf{M}(d_x d_y) \log d_x + \mathsf{M}(d_y) \log d_y)$.

## 7.5. Multiple slope factorization

We are now in a position to generalize Algorithm 7.2 to the case when the bivariate Hensel lifting is done using the multiple slope method from the previous subsection. The general algorithm is fairly technical, so we found it more instructive to illustrate the main ideas on an example. Let

$$\begin{aligned} F &= P\,Q \\ P &= y + z + x + (y + z)\,x^2 \\ Q &= 2x + z + x + (y + 2z)\,x^2. \end{aligned}$$

For generic $\alpha_2, \alpha_3 \in \mathbb{K}^{\neq}$, consider $\tilde{P} = P(x, \alpha_2 t, \alpha_3 t)$, $\tilde{Q} = Q(x, \alpha_2 t, \alpha_3 t)$, and $\tilde{F} = \tilde{P}\,\tilde{Q}$. Then

$$\begin{aligned} \operatorname{supp} \tilde{P} &= \{(0,1), (1,0), (2,1)\} \\ \operatorname{supp} \tilde{Q} &= \{(0,1), (1,0), (2,1)\} \\ \operatorname{supp} \tilde{F} &= \{(0,2), (1,1), (2,0), (2,2), (3,1), (4,2)\}, \end{aligned}$$

whence $\operatorname{hull} \tilde{F}$ is the closed triangle with vertices $(0,2)$, $(2,0)$, and $(4,2)$. Its lower boundary consists of the edge from $(0,2)$ to $(2,0)$ and the edge from $(2,0)$ to $(4,2)$, which correspond to the weights $w = (1,1,1)$ and $w' = (-1,1,1)$ respectively. The generalized algorithm starts with the recursive factorizations of $\mathrm{tp}_w F$ and $\mathrm{tp}_{w'} F$, which yields

$$\begin{aligned} \mathrm{tp}_w F &= (\mathrm{tp}_w P)\,(\mathrm{tp}_w Q) \\ \mathrm{tp}_w P &= y + z + x \\ \mathrm{tp}_w Q &= 2y + z + x \end{aligned} \qquad \begin{aligned} \mathrm{tp}_{w'} F &= (\mathrm{tp}_{w'} P)\,(\mathrm{tp}_{w'} Q) \\ \mathrm{tp}_{w'} P &= x + (y + z)\,x^2 \\ \mathrm{tp}_{w'} Q &= x + (y + 2z)\,x^2. \end{aligned}$$

However, these factorizations do not tell us from which factors of $F$ they originate: the factorization of $F$ could have been of the form $F=P'Q'$, with $\operatorname{tp}_w P'=\operatorname{tp}_w P$ and $\operatorname{tp}_{w'} P'=\operatorname{tp}_{w'} Q$. In order to obtain the correct matches, the next idea is to consider the bivariate factorization of $\tilde{F}$ for some sufficiently generic values of $\alpha_2$ and $\alpha_3$. For instance, for $\alpha_2=2$ and $\alpha_3=3$, we obtain

$$\begin{array}{rclcrclcrcl}\tilde{P} & = & 5t+x+5tx^2 & \quad & \widetilde{\operatorname{tp}_w P} & = & 5t+x & \quad & \widetilde{\operatorname{tp}_{w'} P} & = & x+5tx^2\\ \tilde{Q} & = & 7t+x+8tx^2 & & \widetilde{\operatorname{tp}_w Q} & = & 7t+x & & \widetilde{\operatorname{tp}_{w'} Q} & = & x+8tx^2.\end{array}$$

From these computations, we deduce that the factor $y+z+x$ of $\operatorname{tp}_w F$ comes from the same factor of $F$ as the factor $x+(y+z)x^2$ of $\operatorname{tp}_{w'} F$. We say that we have managed to *match* the factors $y+z+x$ and $x+(y+z)x^2$ of the different slopes. In other words, if $A$ is a non-trivial factor of $F$, then, up to constant scaling, we now know that either $(\operatorname{tp}_w A, \operatorname{tp}_{w'} A)=(\operatorname{tp}_w P, \operatorname{tp}_{w'} P)$ or $(\operatorname{tp}_w A, \operatorname{tp}_{w'} A)=(\operatorname{tp}_w Q, \operatorname{tp}_{w'} Q)$.

For any $i\in\mathbb{N}$ and $A\in\mathbb{K}[x,y,z]$, let $A^{[i]}(x,t)=A(x,\alpha_2^i t,\alpha_3^i t)$. Having completed the above matching, we may now use the lifting algorithm from Section 7.4 to deduce $P^{[i]}$ and $Q^{[i]}$ from $F^{[i]}$, $(\operatorname{tp}_w P)^{[i]}$, $(\operatorname{tp}_{w'} P)^{[i]}$, $(\operatorname{tp}_w Q)^{[i]}$, and $(\operatorname{tp}_{w'} Q)^{[i]}$. Doing this for sufficiently many $i$, we may finally recover $P$ and $Q$ using sparse interpolation.

## 7.6. A torture example

The approach from Section 7.5 is fairly general. However, as we will show now, it is possible to construct pathological examples that can still not be treated with this method.

**Example 7.5.** Consider a polynomial $P\in\mathbb{K}[x_1,\ldots,x_n]$ and a monomial $M=x_1^{e_1}\cdots x_n^{e_n}$ such that $(e_1,\ldots,e_n)$ lies in the interior of the Newton polytope hull $P$. For instance, we may take

$$\begin{aligned} P &:= x_1x_2+x_1x_3+x_2x_3+x_1^2x_2^2x_3+x_1^2x_2x_3^2+x_1x_2^2x_3^2\\ M &:= x_1x_2x_3.\end{aligned}$$

Now consider

$$F = (P+\alpha_1 M)\cdots(P+\alpha_\ell M),$$

for pairwise distinct $\alpha_1,\ldots,\alpha_\ell\in\mathbb{K}$. This polynomial cannot be factored using the techniques from this section, so far.

## 7.7. Factor matching through homotopies

In fact, it is not so much the factorization of the bivariate $F^{[i]}$ polynomials that is a problem: instead of Hensel lifting, we may very well rely on Theorem 4.7 (this only increases the exponent in $d$ in Theorems 7.1 and 7.2, which is subdominant anyway if $s\ll\bar{s}$). The true problem is matching corresponding factors of $F^{[i]}$ for different $i$, especially in the most difficult cases like Example 7.5. We conclude this section with an approach that completely solves this problem, modulo a polynomial overhead in $d$.

Let $\alpha_1,\ldots,\alpha_n,\beta_1,\ldots,\beta_n,\gamma_1,\ldots,\gamma_n$ be random elements of $\mathbb{K}^{\neq}$ and let $t,u,\lambda$ be new indeterminates. We define

$$\begin{aligned}\hat{F}(x_1,\ldots,x_n,t,u,\lambda) := F(&((1-\lambda)+\alpha_1\lambda)(t+\beta_1 u+\gamma_1)x_1,\ldots,\\ &((1-\lambda)+\alpha_n\lambda)(t+\beta_n u+\gamma_n)x_n). \end{aligned} \tag{7.2}$$

For $i=1,2,\ldots,$ we also define

$$F^{\langle i\rangle}(t,u,\lambda) \ := \ \hat{F}(\alpha_1^i,\ldots,\alpha_n^i,t,u,\lambda).$$

If $F$ is irreducible, then $F^{\langle i\rangle}$ is irreducible with high probability, by Theorem 1.1, and so are $F^{\langle i\rangle}(t,u,0)$ and $F^{\langle i\rangle}(t,u,1)$. For general $F$, this also means that the factors of $F^{\langle i\rangle}(t,u,\lambda)$, $F^{\langle i\rangle}(t,u,0)$, and $F^{\langle i\rangle}(t,u,1)$ are in effective one to one correspondence, with high probability. Moreover, by construction,

$$F^{\langle i+1\rangle}(t,u,0) \ = \ F^{\langle i\rangle}(t,u,1).$$

Using this relation, we may therefore match corresponding factors of $F^{\langle i\rangle}$ and $F^{\langle i+1\rangle}$ with high probability.

Having solved the matching problem, we still need to ensure that the factorizations of the $F^{\langle i\rangle}$ are normalized in a suitable way such that we can recover the factors of $F$ using sparse interpolation. For this, let $w$ be a regularizing weight for $F$ with $e:=\operatorname{ec}_w F\leqslant d^2$. Let $\xi$ be yet another indeterminate and consider

$$\begin{aligned}\tilde{F}(x_1,\ldots,x_n,t,u,\lambda,\xi) \ &:= \ F(((1-\lambda)+\alpha_1\lambda)\,(t+\beta_1 u+\gamma_1)\,x_1\,\xi^{w_1},\ldots,\\ &\qquad\quad ((1-\lambda)+\alpha_n\lambda)\,(t+\beta_n u+\gamma_n)\,x_n\,\xi^{w_n})\\ \tilde{F}^{\langle 1\rangle}(t,u,\lambda,\xi) \ &:= \ \tilde{F}(\alpha_1,\ldots,\alpha_n,t,u,\lambda).\end{aligned}$$

By construction, any factor $\Phi$ of $\tilde{F}^{\langle 1\rangle}$ has leading coefficient $c\,M^{\langle 1\rangle}\,t^\tau\,u^\upsilon\,\lambda^\sigma$ with respect to $\xi$, for some $\tau,\upsilon,\sigma\in\mathbb{N}$, $c\in\mathbb{K}^{\neq}$, and some monomial $M\in x_1^{\mathbb{N}}\cdots x_n^{\mathbb{N}}$. Thus, any such factor $\Phi$ can be normalized by dividing out the constant $c\,M^{\langle 1\rangle}$ in $\mathbb{K}^{\neq}$. Now consider the unique factorization of $\tilde{F}^{\langle 1\rangle}$ into a product of normalized factors times a non-zero scalar in $\mathbb{K}^{\neq}$. This yields a factorization of $F^{\langle 1\rangle}$ by setting $\xi=1$.

The factorization of $F^{\langle 1\rangle}$ can be lifted to the factorization of $F^{\langle 2\rangle}$, which is, in turn, lifted to the factorization of $F^{\langle 3\rangle}$, and so on. An irreducible factorization of $\hat{F}$ is recovered via interpolation from the factorizations of $F^{\langle 1\rangle},\ldots,F^{\langle m\rangle}$ for sufficiently large $m$. Finally, to obtain a factorization of $F$ from a factorization of $\hat{F}$, we can set $t:=u:=\lambda:=0$, and apply the map $x_i\mapsto(\gamma_i)^{-1}x_i$.

This leads to the following algorithm:

**Algorithm 7.3**
**Input:** $F\in\mathbb{K}[x_1,\ldots,x_n]\setminus\mathbb{K}$
**Output:** an irreducible factorization $F=c\,P_1^{\nu_1}\cdots P_\ell^{\nu_\ell}$ of $F$

Compute $\hat{F}$
Let $w$ be a regularizing weight for $F$
Compute an irreducible factorization $F^{\langle 1\rangle}=c\,(P_1^{\langle 1\rangle})^{\nu_1}\cdots(P_\ell^{\langle 1\rangle})^{\nu_\ell}$, normalized as above
For $m=1,2,4,8,\ldots$ do
  Compute $F^{\langle i\rangle}$ for $i=\lfloor m/2\rfloor+1,\ldots,m$
  For $i=\lfloor m/2\rfloor+1,\ldots,m$ do
    Hensel lift $F^{\langle i-1\rangle}(t,u,1)=c\,(P_1^{\langle i-1\rangle}(t,u,1))^{\nu_1}\cdots(P_\ell^{\langle i-1\rangle}(t,u,1))^{\nu_\ell}$
      into an irreducible factorization $F^{\langle i\rangle}=c\,(P_1^{\langle i\rangle})^{\nu_1}\cdots(P_\ell^{\langle i\rangle})^{\nu_\ell}$
    Deduce $P_1^{\langle i\rangle},\ldots,P_\ell^{\langle i\rangle}$ from $(P_1^{\langle i\rangle})^{\nu_1},\ldots,(P_\ell^{\langle i\rangle})^{\nu_\ell}$ via root extraction
  Try to determine $P_1,\ldots,P_\ell$ from $P_1^{\langle i\rangle},\ldots,P_\ell^{\langle i\rangle}$ $(i=1,\ldots,m)$ using sparse interpolation
    If successful, then set $t:=u:=\lambda:=0$, $x_i:=(\gamma_i)^{-1}x_i$ and return $c\,P_1^{\nu_1}\cdots P_\ell^{\nu_\ell}$

THEOREM 7.6. *Let* $s := \max(s_{P_1},\ldots,s_{P_\ell})$, $\bar{s} := \max(s,s_F)$, $d := \deg F$, *and* $e := \mathrm{ec}_w F \leqslant d^2$. *Assume that* char $\mathbb{K} = 0$ *or* char $\mathbb{K} > 2d^2$. *Then Algorithm 7.3 runs in time*

$$O(\mathsf{S}(d^3\bar{s}) + \mathsf{M}(d^3)\, s\log d) + \tilde{O}(e^5) + \mathsf{F}(e+3d)$$

*and succeeds with probability at least* $1 - \left(360\, d^2 \binom{s}{2} + 2\, s\, d^3 + 3\,(e+3d)^2 + d\right)/N$.

**Proof.** We factor $\tilde{F}^{\langle 1\rangle}$ as a dense polynomial in four variables of degree $\leqslant e+3d$ (after division by a suitable power of $\xi$) using the algorithm from [72, Proposition 5]. This requires $\tilde{O}(e^5) + \mathsf{F}(e+3d)$ operations in $\mathbb{K}$ and the probability of success is at least $1-3\,(e+3d)^2/N$.

Let $\tilde{c}\,\tilde{P}_1^{\nu_1}\cdots\tilde{P}_\ell^{\nu_\ell}$ be an irreducible factorization of $F$. By our assumption on the characteristic, the factors $\tilde{P}_1,\ldots,\tilde{P}_\ell$ are all separable. We will apply [75, Théorème 6.9] for each of the points in our geometric progression. By using Corollary 2.2 instead of the Schwartz-Zippel lemma in the proof of [75, Théorème 6.9], we deduce that the specializations $\tilde{P}_j^{\langle i\rangle}(t,u,0)$, $\tilde{P}_j^{\langle i\rangle}(t,u,1)$, and thus $\tilde{P}_j^{\langle i\rangle}$ are irreducible for $i=1,\ldots,m$ and $j=1,\ldots,\ell$, with probability at least $1-10\,(3d)^2\binom{m}{2}/N \geqslant 1-360\,d^2\binom{s}{2}/N$. Under this assumption, and modulo reordering the factors, the computed $P_j^{\langle i\rangle}$ are of the form $P_j^{\langle i\rangle} = c_j\,\tilde{P}_j^{\langle i\rangle}$ for suitable scaling factors $c_1,\ldots,c_\ell \in \mathbb{K}^{\neq}$ that do not depend on $i$. The check whether the computed factorization of $F$ is correct reports a false positive with probability at most $d/N$, by Remark 2.5 or the Schwarz-Zippel lemma.

Let us now analyze the complexity. We first observe that $s_{\hat{F}} \leqslant O(d^3 s_F)$, since $s_{\hat{F}} = O(d^3)$ when $\hat{F}$ consists of a single monomial of total degree $\leqslant d$. Using [2, Theorem 5] in a recursive manner, we may compute $\hat{F}$ from $F$ in time $O(\mathsf{M}(d^3)\,s_F) = O(\mathsf{S}(d^3 s_F))$. We saw above that the factorization of $F^{\langle 1\rangle}$ requires at most $\tilde{O}(e^5) + \mathsf{F}(e+3d)$ operations in $\mathbb{K}$. The computation of the specializations $F^{\langle i\rangle}$ for $i=1,\ldots,m$ requires $O(\mathsf{S}(s)\,d^3 m/s) = O(\mathsf{S}(d^3\bar{s}))$ further operations. The Hensel lifting of the $P_j^{\langle i-1\rangle}$ can be done in time $O(\mathsf{M}(d^3)\,s\log d)$ using Proposition 4.4 and evaluation-interpolation in the remaining variable. The $m-1$ Hensel lifts succeed with probability at least $1-2\,s\,d^3/N$. Adding up, we obtain the desired complexity bound, as well as the claimed bound for the probability of success. □

**Remark 7.7.** Due to the $O(d^3)$ overhead, Algorithm 7.3 is mainly of theoretical interest. It is plausible that (7.2) can be replaced by a less, but still sufficiently, generic formula. This would reduce the overhead in $d$. In practice, one may use Algorithm 7.3 as a last resort, in the unlikely case that all other strategies from Sections 6 and 7 fail.